\UseRawInputEncoding
\documentclass[fleqn,usenatbib]{mnras}

\usepackage{newtxtext,newtxmath}

\usepackage[T1]{fontenc}
\usepackage{ae,aecompl}
\usepackage{comment}

\usepackage{graphicx}	
\usepackage{amsmath}	
\usepackage{amssymb}	
\usepackage{placeins}

\title[Impact of the ISM magnetic field on GRB afterglow polarization]{Impact of the ISM magnetic field on GRB afterglow polarization}

\author[Teboul \& Shaviv ]{
O. Teboul,$^{1}$\thanks{E-mail: odelia.teboul1@mail.huji.ac.il}
N. J. Shaviv,$^{1}$
\\
$^{1}$The Racah Institute of Physics, The Hebrew University of Jerusalem, Jerusalem 91904, Israel\\
}

\date{Accepted 16 August 2021. Received 30 July 2021; in original form 24 August 2020}

\pubyear{2020}

\begin{document}
\label{firstpage}
\pagerange{\pageref{firstpage}--\pageref{lastpage}}
\maketitle


\begin{abstract}
Linear polarization has been measured in several GRB afterglows. After a few days, polarization arises from the forward shock emission which depends on the post-shock magnetic field. The latter can originate both from compression of existing fields, here the ISM magnetic field, and from shock generated instabilities. For short GRBs, previous modelling of the polarization arising from the forward shock considered a random field fully or partially confined to the shock plane. However, the ISM magnetic field likely consists of both random and ordered components. Here we study the impact of a more realistic magnetic field having both ordered and random components. We present our semi-analytical model and compute polarization curves arising for different magnetic field configurations. We find that the presence of an ordered component, even significantly weaker than the random one, has distinct signatures that could be detectable. In the presence of an ordered component not in the observer plane, we show that: i) for an observer inside the jet, the polarization angle $\theta_p$ either remains constant during all the afterglow phase or exhibits variations smaller than the 90\textdegree swing expected from a random component solely, ii) for an off-axis observer, the polarization angle evolves from $\theta_p^{\max}$, before the jet break to its opposite after the jet break. We also find that the upper limit polarization for GRB170817 requires a random field not fully confined to the shock plane and is compatible with an ordered component as large as half the random one. 

\end{abstract}




\section{Introduction}
\label{sec:intro}
Since the first successful Gamma Ray Burst (GRB) polarization detections \citep{Covino99,Wijers}, GRB polarization has been detected in numerous GRBs , see for reviews\citep{ Covino04,Covino16}. Polarization is a unique tool that can shed light on GRB scenarios. Indeed, the polarization depends on  parameters such as the magnetic field configuration which are hardly distinguishable using flux detections only. {\color{black}The last few years} have been notably rich for GRB polarization detections. Polarization was detected for the prompt emission \citep{zhang} and both the early \citep{laskar,jordana} and late afterglow \citep{corsi}. The early and late afterglows are due to two different shocks namely reverse and forward shocks. When the GRB jets interact with their ambient medium two shocks are generated: a relativistic forward shock (FS) that travels into the ambient medium and a short-lived reverse shock (RS) which propagates back into the jet \citep{SP:99,kobayashi}. While the late afterglow is due to the forward shock, the early afterglow can arise from just the reverse shock or a combination of both shocks \cite[as in][]{jordana}. 

Forward shock emission is arising {\color{black} from synchrotron emission of radiating electrons} and depends on the local magnetic field behind the forward shock. The magnetic field behind the shock can originate both from compression of an existing magnetic field (the ISM magnetic field, \citealt{laing}), and from shock generated two-stream instabilities such as the Weibel instability \citep{medvedev}. The compressed ISM magnetic field has long been regarded as too weak. However, recent studies have found that the typical circumburst density for short GRBs is $n\approx 10^{-3} cm^{-3}$ \citep{fong}, while the typical fraction of post-shock magnetic energy is $\epsilon_B\approx 10^{-3} $ \citep{santana}, with some extreme cases having $\epsilon_B \sim 10^{-6}$ (e.g \citealt{barniol,he,kumar}). Moreover, the fraction of magnetic energy is defined by $\epsilon_B = U_B /e_\mathrm{th}$ with $U_B = B^2/8 \pi$ with $e_\mathrm{th}\approx 4 \Gamma^2 n m_p c^2 $. Thus, the {\color{black}comoving} magnetic field behind the shock is given by: 
\begin{equation}
  B=(32\pi n m_p \epsilon_B)^{1/2} \Gamma c .
\end{equation}
 {\color{black}with $\Gamma$ the Lorentz factor of the shock.} 
We have $B\approx 10^{-2} (\epsilon_B/ 10^{-3}) (n/ 10^{-3} cm^{-3}) (\Gamma/10^2) \:$G for these typical values. On the other hand, the compressed ISM magnetic field has a strength $B_\mathrm{comp} = 4 \Gamma B_\mathrm{ISM} $ which for an ISM magnetic field of a few $\mu G$ gives $B_\mathrm{comp} \approx 10^{-3} (\Gamma/10^2) \:$G. The compressed ISM magnetic field is therefore not negligible and can even play a key role in events with low density and low $\epsilon_B $ such as GRB170817. 

Observations have shown that the  ISM magnetic field typically consists of three components: a large-scale coherent component, a small-scale random or turbulent component, and a striated component that changes direction stochastically on small scales but whose orientation remains aligned over large scales (\citealt{boulanger} and references therein).  Moreover, Planck measurements estimate the ratio between the typical strengths of the turbulent and large scale components of the field to be 0.8 \citep{planckadam}. Therefore, we can expect the large scale ordered component to be dominant in the pre-shocked ISM, and compression of such an ISM magnetic field will result in a magnetic field behind the shock with some ordered component. 

So far, previous models for GRB afterglow polarization thoroughly considered a random magnetic field confined to the shock plane \citep{Ghisellini:99,Sari:99,Granot:03,rossi,Gill:18} or an anisotropic random magnetic field \citep{Sari:99,Granot:03,Gill:18}. \cite{Granot:03} noted that progenitor stars have magnetized winds which could give rise to an ordered component, and calculated the polarization under the limits that the ordered and random components originate from different fluid elements, that the ordered component only gave the maximum synchrotron polarization, and that the observer was inside the jet. 

In this work, we study the impact of different magnetic field configurations including the one consisting of both ordered and random components, in the general case where both components can be present in the same fluid element, and for an observer both outside and inside the jet. We begin in \S\ref{sec:pointlike} with the derivation of the Stokes parameters for a point like region arising from the different magnetic field configurations. Next, we present in \S\ref{sec:afterglows} our semi-analytic calculations of the jet flux and its evolution, which we use in \S\ref{sec:jetpolarization} to calculate polarization curves for the different magnetic field configurations. In \S\ref{sec:GRB170817} we apply our results to the observed upper limit on the polarization of GRB170817 \citep{corsi}. GRB170817 was the first electromagnetic counterpart to a gravitational waves event and was therefore extensively monitored. Hence, parameters of both the jet and the ambient medium are better constrained than usual, making it an ideal case to study the magnetic field configuration behind the shock. Finally in \S\ref{sec:conclusions} we discuss our results. 

\section{Polarization of a point like region}
\label{sec:pointlike}

We begin by calculating the polarization arising from a point like 
region, i.e. a region where the direction towards the observer is constant and over which the magnetic field is entangled. In order to obtain the polarization of such a region, we need to integrate over all the possible orientations of the magnetic field. 

\begin{figure}
	\centerline{\includegraphics[width=90mm]{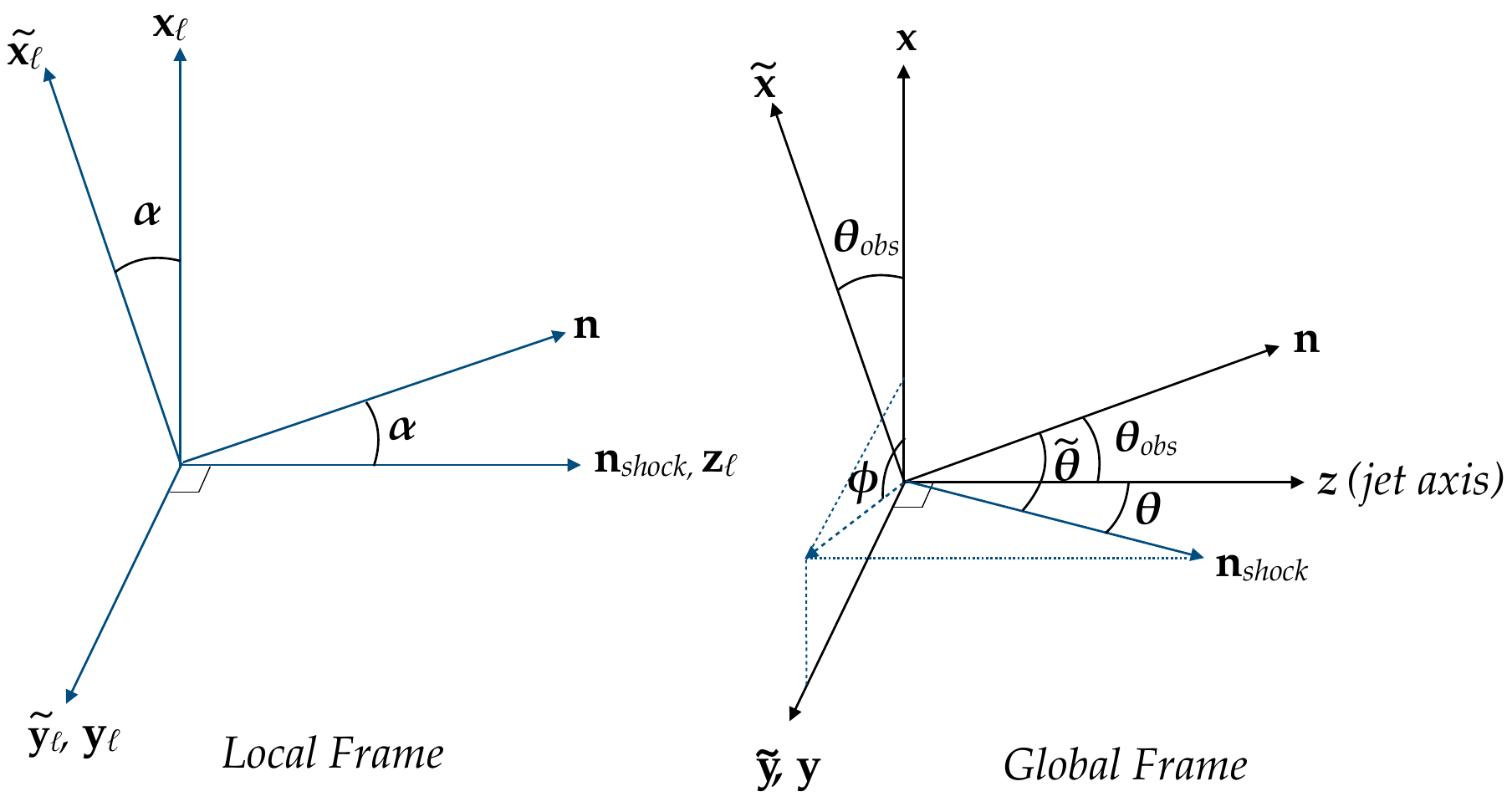}}
    \caption{Global and local frames.}
   \label{fig:frames}
\end{figure}

At each time, the jet is moving towards a preferred direction, the $\bf{z}$ axis. The observer is then chosen to lie in the (x,z) plane (see fig.\ \ref{fig:frames}). Each fluid element of the jet has a slightly different velocity vector  ${\bf n}_\mathrm{shock} $ which depends on fluid element coordinates $(r,\theta, \phi)$ and on time (see fig.\ \ref{fig:frames}). 

Let us consider a point like region whose velocity vector is ${\bf n}_\mathrm{shock} $, we choose this direction to be the $\bf{z}_\ell$ axis see fig.\ \ref{fig:frames}, and work in the rest of this section in this local coordinate system. If the observer direction is  $\bf{n} $ then  the polarization  vector is in the $\bf{n} \times \bf{B} $ direction, and only the magnetic field perpendicular to the observer is contributing to the polarization. {\color{black} We assume that the radiating electrons have a power law distribution of index $ 2\alpha +1 $ giving rise to synchrotron emission proportional to some power of the magnetic field $B^{\alpha +1}$. For $\alpha =1$   a simple solution exists. In this case, the Stokes parameters are given by:}
   
\begin{equation}
\begin{aligned}
\frac{Q}{I} =  \Pi_{max} \frac{\left\langle[(\bold{n} \times \bold{B} ) \cdot \bold{\tilde{x}_\ell}]^{2}\right\rangle-\left\langle[(\bold{n} \times \bold{B} ) \cdot \bold{\tilde{y}_\ell}]^{2}\right\rangle}{\left\langle[(\bold{n} \times \bold{B} ) \cdot  \bold{\tilde{x}_\ell}]^{2}\right\rangle+\left\langle[(\bold{n} \times \bold{B} ) \cdot \bold{\tilde{y}_\ell}]^{2}\right\rangle}
\end{aligned}
\end{equation}

\begin{equation}
\color{black}
\begin{aligned}
\frac{U}{I} &=  \Pi_{max} \frac{\left\langle[(\bold{n} \times \bold{B} ) \cdot  \bold{e_{a,\ell}}]^{2}\right\rangle-\left\langle[(\bold{n} \times \bold{B} ) \cdot \bold{e_{b,\ell}}]^{2}\right\rangle}
{\left\langle[(\bold{n} \times \bold{B} ) \cdot  \bold{\tilde{x}_\ell}]^{2}\right\rangle+\left\langle[(\bold{n} \times \bold{B} ) \cdot \bold{\tilde{y}_\ell}]^{2}\right\rangle}
\end{aligned}
\end{equation}

where {\color{black} $\Pi_{max}=({3\alpha +3})/({3\alpha + 5})$ }, $\bf{n}$ is the observer direction, $\bold{\tilde{x}_\ell}$ and $\bold{\tilde{y}_\ell}$ are the local axis of the plane of polarization and, $\bold{e_{a,\ell}}$ and $\bold{e_b,\ell }$ are the axis making a 45\textdegree\ angle  with $(\bold{\tilde{x}_\ell},\bold{\tilde{y}_\ell} )$.
The polarization of a point like region is given by:
\begin{equation}
\Pi_p = \frac{\sqrt{Q^2+U^2}}{I}.
\end{equation}

\subsection{Random field}
A three dimensional random magnetic field $\bf{B}$ is following a  multivariate Gaussian law: 
\begin{equation}
{\bf B} \sim \mathcal{N}(0,\,\Sigma) \quad 
     \mathrm{with} \quad   \Sigma =  \mathrm{Diag} \left( \sigma_\perp^2, \sigma_\perp^2,
\sigma_\parallel^2 \right).
\end{equation}
For an isotropic magnetic field $\sigma_\parallel= \sigma_\perp$ and  the polarization vanishes. Whereas, for an anisotropic magnetic field  with an anisotropy factor $b= \sigma_\parallel^2/  \sigma_\perp^2$, we obtain the following polarization for a point like region:
\begin{equation}
 \Pi_p =\frac{Q}{I}={\color{black}\Pi_{max}} \frac{\sin^2 \alpha (1-b)}{ 1+ \cos^2 \alpha +  b \sin^2 \alpha }  ,\qquad  U=0,
  \label{eq:pola_rand_zone}
 \end{equation}
with $\alpha$ the angle between the observer and the point like emitting region.
This is the same expression than \cite{gruzinov} and \cite{Sari:99} \footnote[1]{These analyses considered a positive polarization to be along the $y$ axis, while we consider a positive polarization along the $\tilde{x}$ axis.}.
If the magnetic field is fully confined to the shock plane, i.e. $b = 0$, we recover the 2D isotropic case obtained by \cite{laing}. \footnote[2]{Note that here $\alpha$ is the angle between the jet and the observer while in \cite{laing} $\beta$ is the angle between the observer and the plane of the slab, therefore $\alpha=\frac{\pi}{2}-\beta$.} 

\subsection{Ordered field and  random field}
If the magnetic field before compression is composed of a large scale ordered component ${\bf B}_0$ and a random component ${\bf B}_\mathrm{rand}$, it can be written as 
 \begin{equation}
\textbf{B} = \mu B_\mathrm{rand} \bf{n_0} + \bf{B}_\mathrm{rand}
  ~~\mathrm{where}~~ \mu \equiv \sqrt{ {\left\langle B_0^2 \right\rangle}/{\left\langle B_\mathrm{rand}^2 \right\rangle}}
 \end{equation}
 and $\bf{n_0}$  is the direction of the ordered field before compression.

Behind the shock, the magnetic field follows a multivariate Gaussian distribution 
\begin{equation}
\textbf{B} \sim \mathcal{N}(B_{0,\perp},\,\Sigma)\,,
  \end{equation}
with $\bf{B_{0,\perp}}$ the projection of $\bold{B_0}$ in the plane perpendicular to the shock.
In the most general case of a magnetic field composed of an ordered field and an anisotropic random field, we obtain the following Stokes parameters:
\begin{equation}
I = 1 +   \cos^2 \alpha+ b \sin^2 \alpha + \mu^2\left[n_{0,y}^2 +  n_{0,x}^2 \cos^2 \alpha \right] ,
  \label{eq:I} 
\end{equation}\vskip -6mm
\begin{equation}
 Q =\Pi_\mathrm{max}  \left(\sin^2 \alpha (1-b) +\mu^2\left[n_{0,y}^2 -  n_{0,x}^2 \cos^2 \alpha \right]\right),
  \label{eq:Q} 
\end{equation}\vskip -6mm
\begin{equation}
U = \Pi_\mathrm{max} \mu^2 \left[-2 n_{0,y} n_{0,x} \cos \alpha \right],
  \label{eq:U}
\end{equation}
with $\alpha$ the angle between the observer and the point like emitting region, $b$ the anisotropy factor of the random field, while $n_{0,x}$ and $n_{0,y}$ the projections of the  magnetic field ordered component onto the plane perpendicular to the shock direction.

\section{Afterglow Spectrum and Light Curves}
\label{sec:afterglows}
\subsection{Jet dynamics}
We developed a semi-analytical model for afterglows to compute the light curves. The hydrodynamics is described by a blast wave expanding into a cold medium, assuming an adiabatic flow. The blastwave evolution follows the Blandford and McKee self similar solution \citep{blandford} which gives us the following expressions for the density, Lorentz factor and energy of the shocked fluid:
\begin{equation}
\begin{aligned}
n^{\prime} & =2^{3 / 2} \Gamma n_\mathrm{ext} \chi^{-5 / 4}, \\
 \gamma & =2^{-1 / 2} \Gamma \chi^{-1 / 2}, \\
  e^{\prime}&=2 \Gamma^2  \rho_\mathrm{ext}  c^{2} \chi^{-17 / 12},
\end{aligned}
\end{equation}
{\color{black}where the primed quantities are in the comoving frame}, $\Gamma$  is the Lorentz factor of the shock and $\chi$ is the coordinate of a fluid element:
\begin{equation}
\chi= (1+8 \Gamma^{2})\left(1- \frac{r}{ct}\right)
\end{equation}
{\color{black}with r the radial coordinate and t the lab-frame time.} The initially relativistic blastwave will eventually be decelerated to reach the Newtonian phase when $\Gamma \approx \sqrt{2} $. In the Newtonian phase, the evolution follows the Sedov and Taylor solution \citep{Sedov,Taylor}.
We use the solution for an explosion in a constant density medium $\rho_\mathrm{ext}={\color{black}n_\mathrm{ext}} m_p$, corresponding to an interstellar medium (ISM), with $n$ the number density and $m_p$ the proton mass.
We assume that the magnetic field gets a fixed fraction $\epsilon_B $ of the internal energy everywhere behind the shock, {\color{black} This would be the case if the magnetic field decreases due to adiabatic expansion. Moreover, different assumptions on the evolution and orientation of the magnetic field were shown to have only a small effect on the resulting spectrum \citep{GPS:99} }, cf however  \citep{GG} for an evolving  $\epsilon_B $.  
A constant fraction $\epsilon_e$ of the shock energy goes into the electrons which are considered to acquire a power law distribution of energy, immediately behind the shock $N(\gamma_{\color{black}e}) \propto \gamma_{\color{black}e} ^{-p}$.  

\subsection{Spectrum and light curves}
Synchrotron radiation is the dominant emission mechanism throughout the afterglow arising from the forward shock. Therefore we will neglect Inverse Compton radiation. The spectrum in the comoving frame for slow cooling, the principal phase throughout the afterglow, is given by: 
\begin{equation}
P^{\prime}_{\nu^{\prime}} =   \left\{\begin{array}{ll}
P^{\prime}_{\nu', \max} \left(\nu^{\prime} / \nu_{m}^{\prime}\right)^{1 / 3} & \nu^{\prime}<\nu_{m}^{\prime}<\nu_{c}^{\prime} \\
P^{\prime}_{\nu', \max}\left(\nu^{\prime} / \nu_{m}^{\prime}\right)^{(1-p) / 2} & \nu_{m}^{\prime}<\nu^{\prime}<\nu_{c}^{\prime} \\
P^{\prime}_{\nu', \max}\left(\nu^{\prime} / \nu_{m}^{\prime}\right)^{(1-p) / 2}\left(\nu^{\prime} / \nu_{c}^{\prime}\right)^{-1 / 2} & \nu^{\prime}> \nu_{c}^{\prime},
\end{array}\right.
\end{equation}

with  $P_{\nu^{\prime}}^{\prime}$ the radiated power per unit volume per unit frequency, $\nu_{m}^{\prime}$ the typical synchrotron frequency, and $\nu_{c}^{\prime}$ the cooling  frequency. All the primed quantities pertain to the local rest frame of the fluid. $\nu'= \nu \gamma (1-\beta  \cos \Tilde{\theta}) $ where $\beta c$ is the velocity of the matter emitting the radiation and $\Tilde{\theta}$ is the angle between the direction of the velocity of the matter and the observer in the observer frame.  
$P'_{\nu', \max}, \nu_{c}^{\prime}$ and $\nu_{m}^{\prime}$ are calculated following \cite{GS}. 

The observer flux is calculated by integrating over the contributions from all the shocked region arriving at the same observer time $t_\mathrm{obs}$, following \cite{GPS:99}:
\begin{eqnarray}
F_{\nu}\left(t_{\mathrm{obs}}\right)\hskip -2mm &=& \hskip -2mm\frac{4 R_{l}^{3}(1+z)}{\pi d_{\mathrm{L}}^{2}}
\\ \nonumber && \hskip -2mm\times 
\int_{0}^{2 \pi} \hskip -1mm \mathrm{d} \phi  \int_{1}^{\chi_{\max }} \hskip -1mm\mathrm{d} \chi \int_{0}^{\chi^{-\frac{1}{4}}} \hskip -1mm\mathrm{d} y \frac{\chi y^{10} P_{\nu^{\prime}}^{\prime}\left(y, \chi, t_{\mathrm{obs}}\right)}{\left(1+7 \chi y^{4}\right)^{2}},
\end{eqnarray}
with  $d_{L}$ and $z$ the luminosity distance and cosmological redshift of the source.
The radiated power  $P'_{\nu'}$  is taken at the coordinate time $t=t_{z}+ r \cos \Tilde {\theta} /c  $  where $t_{z} \equiv t_{\mathrm{obs}} /(1+z)$. We take $\chi_{\max }=1+8 \Gamma^{2}$ and $y \equiv {R}/{R_{l}}$ where $R= R(t)$ is the radius of the shock front and $R_l$ \footnote[3]{  {\color{black}$R_l$ is the radius of the point on the shock front, on the line of sight from which a photon reaches the detector at $t_z$ and E is the energy of the blastwave}} is given by:
\begin{equation}
R_{l}=\left[\frac{17 \times 4 E t_{z}}{4 \pi {\color{black}\rho_{ext}} c}\right]^{1 /4}.
\end{equation} 

\begin{figure}
\centerline{\includegraphics[width=70mm]{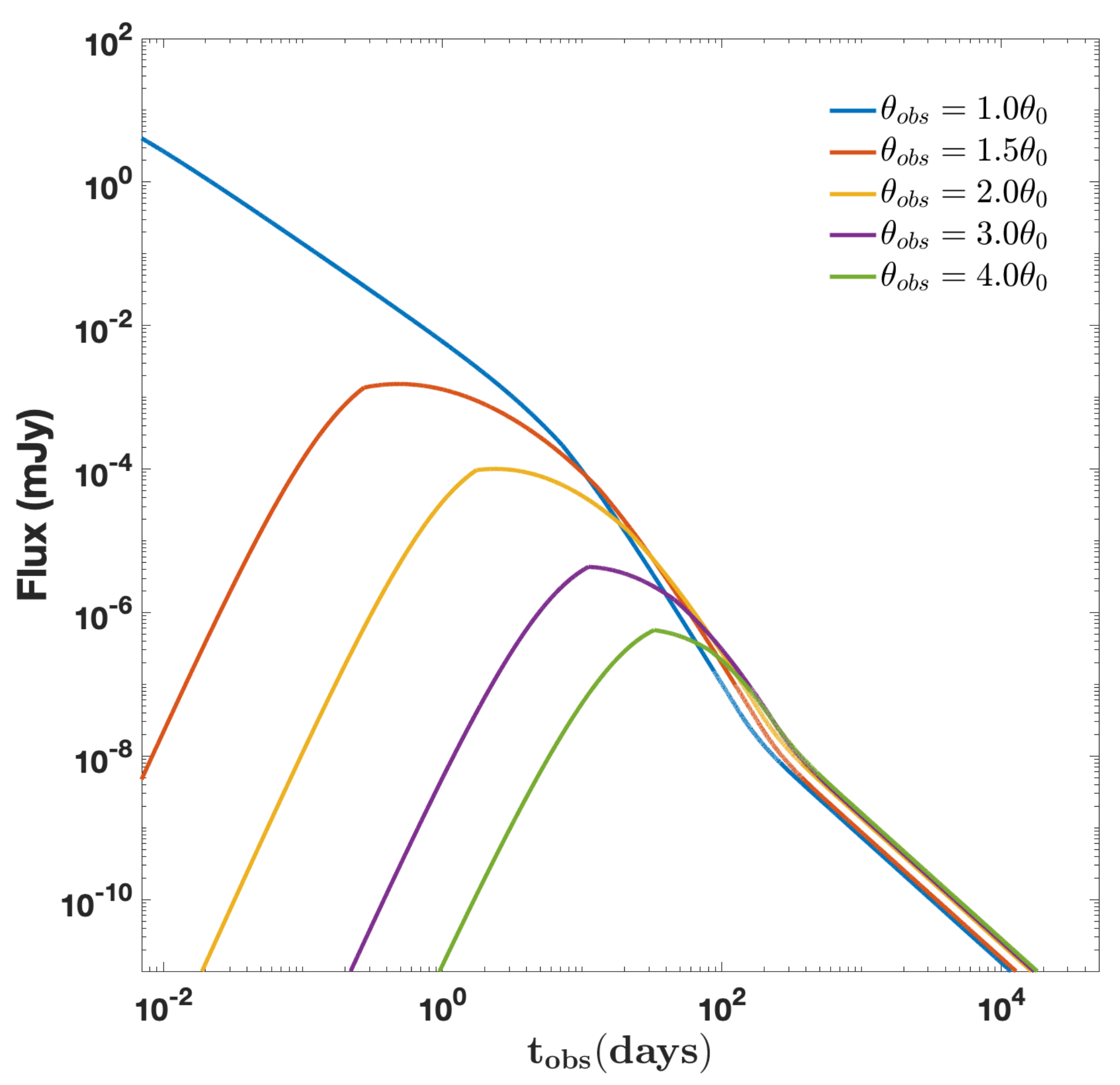}}
    \caption{Light curves for a homogeneous jet for different observer angles. The parameters are $E_\mathrm{iso} =10^{52}$erg, $\theta_0 =10$\textdegree , $n=1 $cm$^{−3},$ $\epsilon_e =0.01,$ $\epsilon_B = 0.005,$ $p=2.5,$ $\nu =7.10^{14 }$Hz.}
   \label{fig:LC}
\end{figure}

{\color{black}The light curves obtained for a homogeneous jet \footnote[4]{$\epsilon(\theta) = \epsilon_0$ for $\theta \leq \theta_0 $} with half opening angle $\theta_0 $ are shown for different observer angles in fig.\ \ref{fig:LC}. }
\FloatBarrier

\section{ Jet Polarization}
\label{sec:jetpolarization}
In \S\ref{sec:pointlike} we derived the Stokes parameters for a point like region arising from different magnetic field configurations. We will now integrate over the total jet emitting region to obtain the jet observed polarization for the different magnetic field configurations.

Each fluid element has a velocity vector ${\bf n}_\mathrm{shock}$ whose coordinates are $(\sin \theta \cos \phi,\sin \theta \sin \phi, \cos \theta) $ with $\theta$ and $\phi$ the spherical coordinates of the fluid element in the global frame of the fluid. The observer lies in the direction ${\bf n} = (\sin \theta_\mathrm{obs}, 0 , \cos \theta_\mathrm{obs})$ and the angle between the velocity vector of a fluid element and the observer $\Tilde{\theta}$  is  $\cos \Tilde{\theta} = {\bf n}_\mathrm{shock} \cdot {\bf n} $. 

In \S\ref{sec:pointlike}, the velocity vector of a point like emitting region was chosen to be in the $\bf{z}_\ell$ direction and the Stokes parameters were obtained in the local frame ($\bf{\Tilde{x}_{\ell}}$,$\bf{y_{\ell}}$)  which depends on ${\bf n}_\mathrm{shock}$. In order to compute all the Stokes parameters in the global frame  ($\bf{\Tilde{x}}$,$\bf{y}$) we calculated the transformations for Stokes parameters. The Stokes parameters $\Tilde{Q}$ and $\Tilde{U}$ in the global frame ($\bf{\Tilde{x}}$,$\bf{y}$) are: 
\begin{equation}
\label{eq:Q2}
\Tilde{Q}=Q \cos (2 \rho) + U \sin (2 \rho), 
\end{equation}\vskip -5mm
\begin{equation}
\label{eq:U2}
\Tilde{U}=-Q \sin (2 \rho) + U  \cos (2 \rho),
\end{equation}
with $Q$ and $U$ the local Stokes parameters and $\rho$ the angle between $\bf{\Tilde{x}_{\ell}}$ and  $\bf{\Tilde{x}}$.

We want to perform the integration in the frame of the fluid and therefore need to take into account the aberration of light. If the angle between the observer and the patch is $\Tilde{\theta}$ in the observer frame then by Lorentz transform of angles it becomes $\theta'$  in the fluid frame:
\begin{equation}
\cos(\theta') = \frac {\cos(\Tilde{\theta}) - \beta }{1-\beta \cos(\Tilde{\theta}) }.
\end{equation}
As the Stokes parameters are additive for incoherent emission, they can be calculated by summing over all the contributions from different fluid elements arriving at the same observer time $t_\mathrm{obs}$. Therefore, the Stokes parameters for the full jet are:
\begin{equation}
\color{black}
{Q_{jet} \over I_{jet}}  =\frac{•\int \tilde{Q}(\theta') \delta ^3 L'_{\nu'}   d\tilde{\Omega}}{\int I(\theta') \delta ^3 L'_{\nu'}   d\tilde{\Omega}},
\end{equation}
\begin{equation}
\color{black}
{U_{jet} \over I_{jet}} =\frac{•\int \tilde{U}(\theta') \delta ^3 L'_{\nu'} d\tilde{\Omega}}{\int I(\theta') \delta ^3 L'_{\nu'}d\tilde{\Omega}},
\end{equation}
with $ L'_{\nu'} $ the luminosity in the fluid frame, $\delta$ the Doppler factor $\delta = [\gamma ( 1- \beta \cos\Tilde{\theta})]^{-1}$, $\tilde{Q}$ and $\tilde{U}$ the Stokes parameters after the change of reference frame.

Finally, the total jet polarization $\Pi$ is given by:
\begin{equation}
\color{black}
\Pi= \frac{\sqrt{Q_{jet}^2+U_{jet}^2}}{I_{jet}}.
\end{equation}

\subsection{Random field confined to the shock plane }
\label{sec:LC random}
\begin{figure}
	\centerline{\includegraphics[width=70mm]{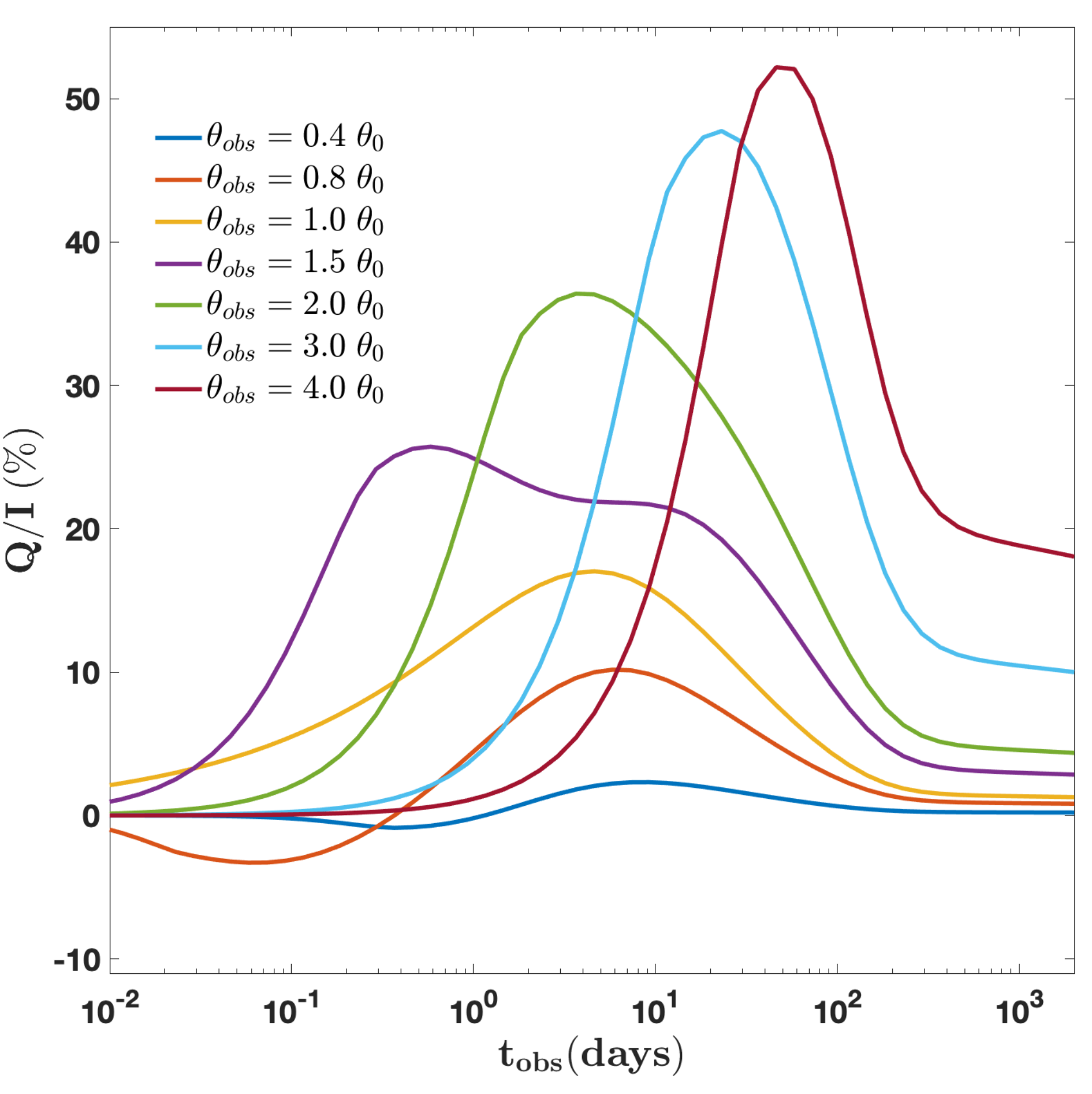}}
    \caption{Polarization curves for a random field confined to the shock plane for different observer angles. The other parameters are the same as in fig. \ref{fig:LC}, ${\color{black}\Pi_{max}} = 60\%$, the curves may have to be rescaled by a factor $< 1 $ , since ${\color{black}\Pi_{max}} = 60\%$ is taken arbitrary.}
   \label{fig:random_figure}
\end{figure}

For a random field fully confined to the shock plane, the polarization of a point like region is given by eq.\ \ref{eq:pola_rand_zone}, with $b=0$.
Integrating over the jet, we obtain the following polarization curves for different observer angles. See fig.\ \ref{fig:random_figure}.

 If $\theta_\mathrm{obs}< \theta_0$, there are two polarization maxima and the direction of polarization rotates by 90\textdegree. However, if $\theta_\mathrm{obs} > \theta_0$, the larger the observer angle is, the later and the stronger does the polarization peaks. We note that the polarization peak is in the same region as the flux peak for an observer outside the jet. These results are similar to those found by \cite{ Ghisellini:99,Granot:03,rossi} \footnote[5]{\cite{Granot:03} and \cite{Ghisellini:99} have computed polarization curves only for an observer inside the jet.}.  

\FloatBarrier

\subsection{Anisotropic Random field}
\begin{figure}
	\centerline{\includegraphics[width=70mm]{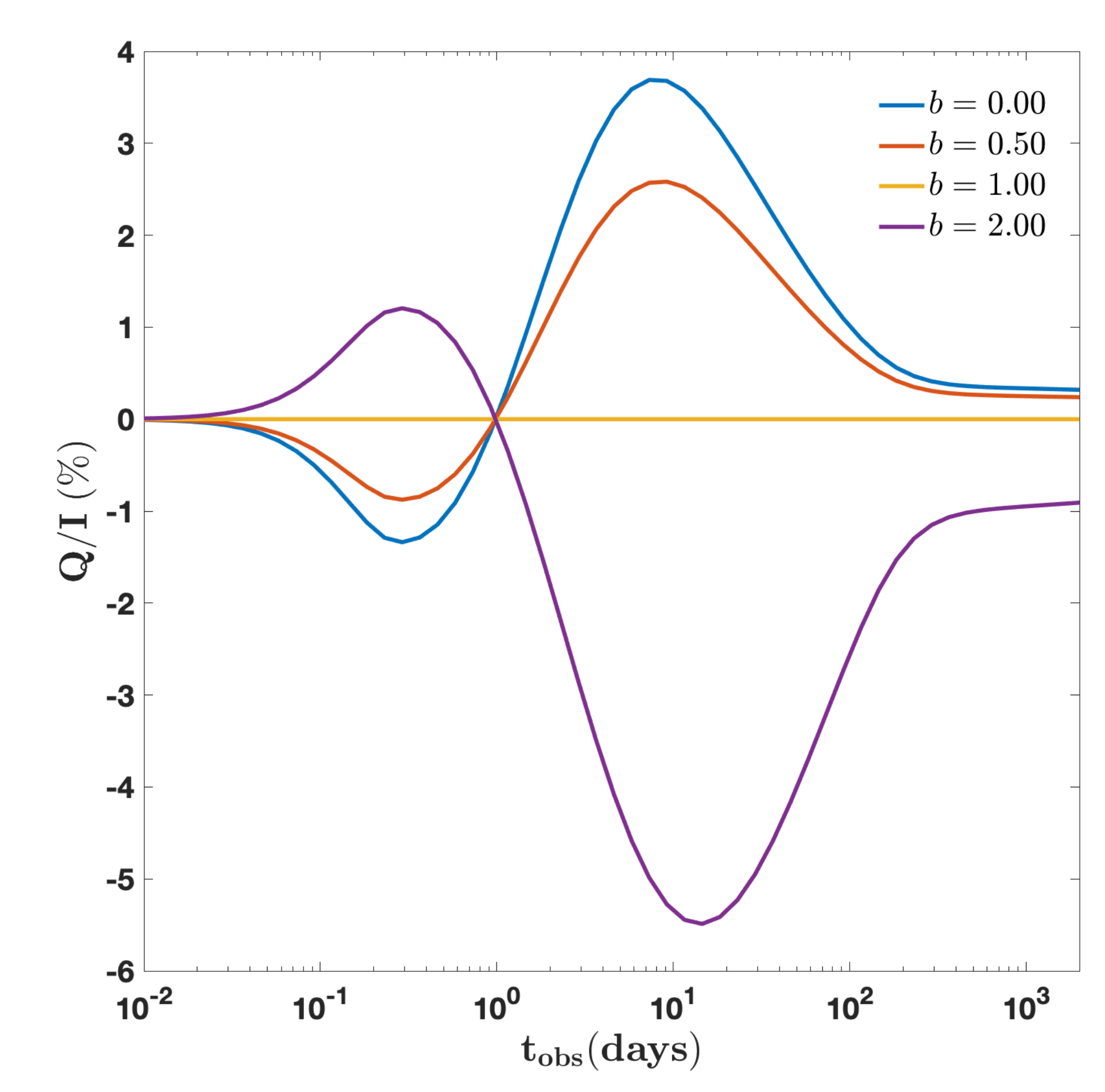}}
    \caption{Polarization curves for an anisotropic random field for different anisotropy factors $b$ at a given observer angle $\theta_\mathrm{obs}=0.5\:\theta_0$. The other parameters are the same as in fig.\ \ref{fig:LC}.}
   \label{fig:anisotropy_figure}
\end{figure}

For an anisotropic random field, the polarization of a point like region is also given by eq  \ref{eq:pola_rand_zone}. In order to isolate the impact of the anisotropy factor $b$, polarization curves are plotted for a given observer angle $\theta_\mathrm{obs} =0.5 \: \theta_0$ and different anisotropy factors in fig.  \ref{fig:anisotropy_figure}.
It can be seen that for all cases there is a 90\textdegree\  swing in the polarization angle. For $b<1$, the polarization is firstly along the $\bf{y}$ direction and then along the $\bf{\tilde{x}}$ direction, while it is the opposite for $b>1$. With solely a random component, there are only two possible directions for the polarization: the $\bf{y}$ and $\bf{\tilde{x}}$ directions. 

\subsection{Ordered field and random field confined to the shock plane}
\label{sec:LC ordered}
As discussed in \S\ref{sec:intro} and \S\ref{sec:pointlike}, the ISM magnetic field before compression most likely consists of both random and ordered components with a larger ordered component \citep{planckadam}. As the random component can also be generated at the shock front, we will consider that behind the shock the magnetic field consists of both ordered and random components, with a larger or an equal random component.

 In the case of a magnetic field with both ordered and random components, the Stokes parameters that we derived are given in eq.\ \ref{eq:I}, \ref{eq:Q} and \ref{eq:U}. An important signature of the presence of an ordered component is that $U \neq  0$, while $U =0 $  with only a random field. We should note that in case of an ordered component in the plane of the observer, there is no break of symmetry and $U$ also vanishes. 
 
\begin{figure}
	\centerline{\includegraphics[width=75mm]{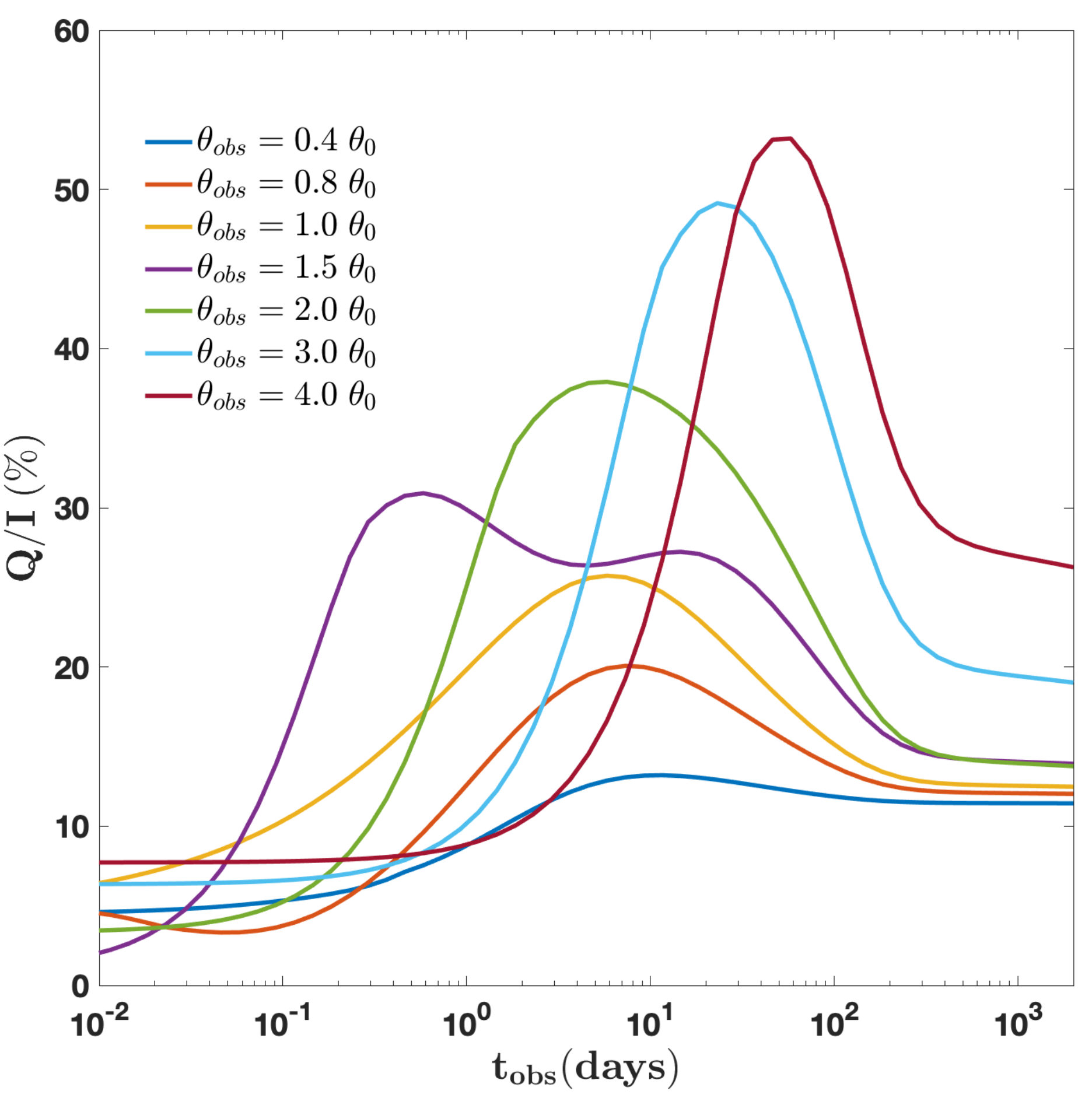}}
    \caption{Q/I evolution for a magnetic field with both ordered and random components confined to the shock plane, with $ \mu=0.5$. The other parameters are the same as in fig. \ref{fig:LC}.}
   \label{fig:Q_ordered}
\end{figure}

\begin{figure}
	\centerline{\includegraphics[width=75mm]{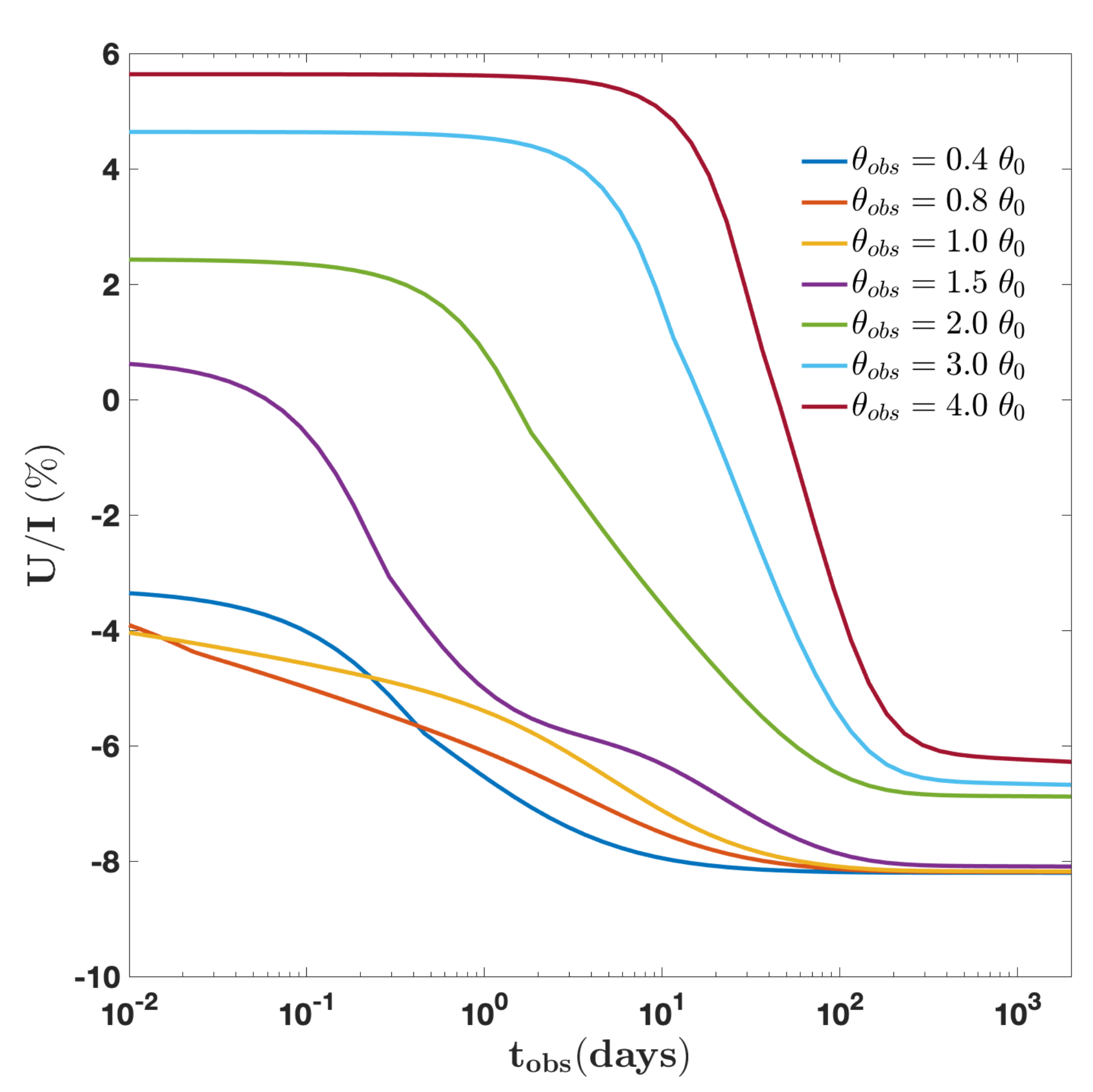}}
    \caption{U/I evolution for a magnetic field with both ordered and random components confined to the shock plane, with $ \mu=0.5$. The other parameters are the same as in fig.\ \ref{fig:LC}.}
   \label{fig:U_ordered}
\end{figure}

 In figs.\ \ref{fig:Q_ordered} and \ref{fig:U_ordered}, the evolution of $Q/I$ and $U/I$ are presented for  an ordered to random component  ratio $\mu =0.5$. The evolution of $Q/I$ resembles the evolution we obtained with solely a random component, with however a higher $Q/I$ at all times for an observer inside the jet and a higher $Q/I$ before and after the polarization peak for an off-axis observer. The $U/I$ evolution depends on whether or not the observer is inside the jet. For an off-axis observer, $U/I$ is first constant until about the time $Q/I$ peaks, then decreases to reach the opposite value. Moreover, the larger the angle is, the higher is the value of $U/I$. For an observer inside the jet, $U/I$ slightly decreases and from the beginning has the opposite value.  
 
 
\begin{figure}
	\centerline{\includegraphics[width=75mm]{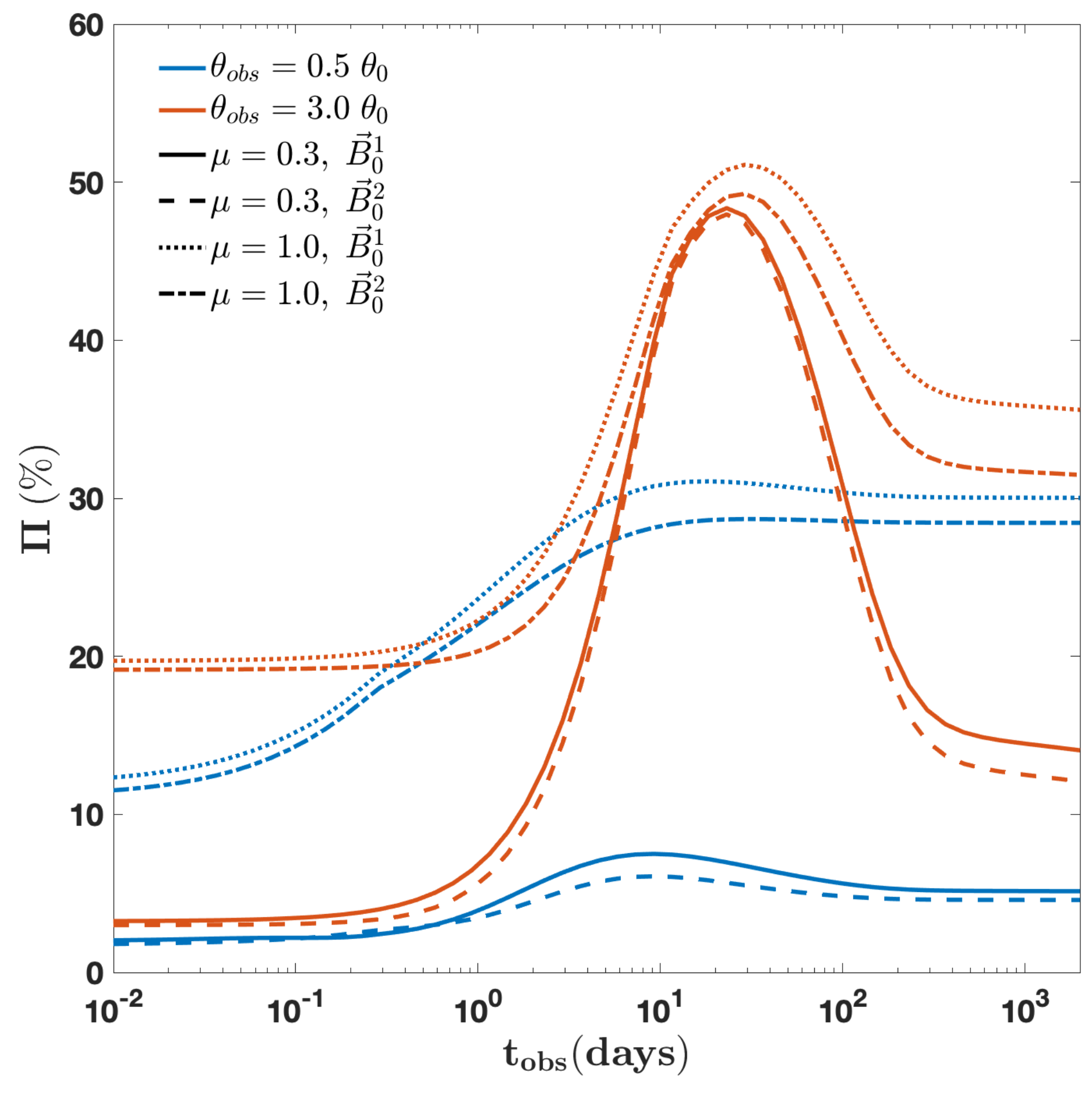}}
    \caption{Polarization curves for two observer angles, two directions of the magnetic field and two ratios of ordered to random component {\color{black}with b=0}. The other parameters are the same as in fig.\ \ref{fig:LC}.}
   \label{fig:ordered_mu}
\end{figure}

In fig.\ref {fig:ordered_mu}, we can see the impact of both  a larger and a weaker ordered component on two different directions of the magnetic field prior to compression. We find that with an ordered component as large as the random one, the polarization is much stronger for both an observer inside and outside the jet. However, for an observer outside the jet, the value of the polarization peak remains roughly constant. 
Therefore, the degree of polarization outside of the peak region is a good indicator of the presence of an ordered component as well as the ratio of ordered to random component.  

\subsection{Ordered field and anisotropic random field}
\begin{figure}
	\centerline{\includegraphics[width=75mm]{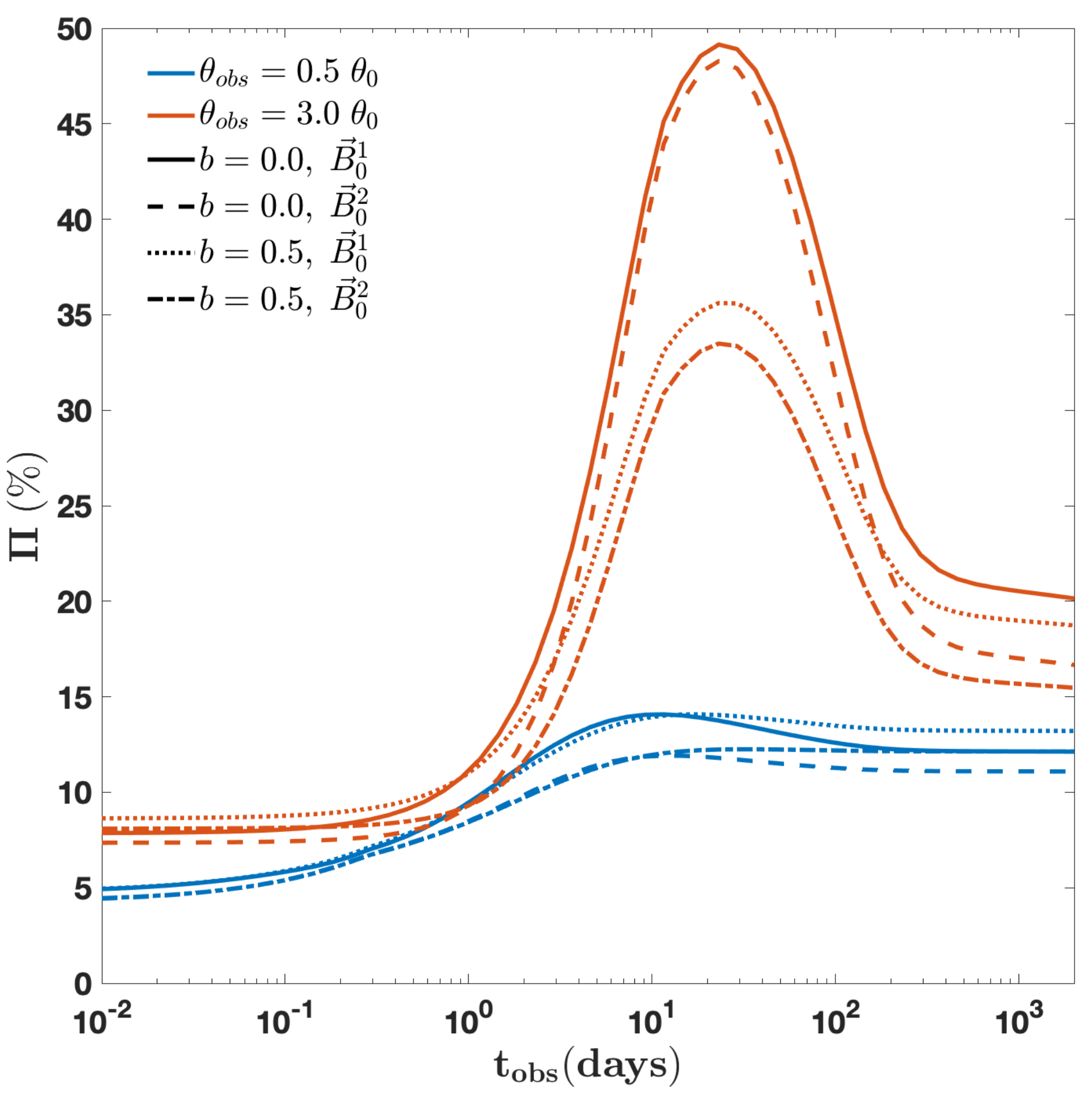}}
    \caption{Polarization curves for two observer angles: inside and outside the jet, two directions of the magnetic field and two anisotropy factors. The other parameters are the same as in fig.\ \ref{fig:LC}.}
   \label{fig:Pi_anisotropy_Bo}
\end{figure}

In this section we consider the more general case of a magnetic field consisting of both an ordered and an anisotropic random component. In order to isolate the impact of the anisotropy in the random component, we present in fig.\ \ref{fig:Pi_anisotropy_Bo} the polarization curves we obtained for the two directions considered previously and two anisotropy factors, with an ordered to random component ratio of $ \mu =0.5$. We can see that  for an observer inside the jet, a different anisotropy factor has a very small impact while for an observer outside the jet, an anisotropic random field strongly reduces the polarization peak.
\label{sec:LC ordered and anisotropic}

\subsection{Polarization angle evolution}
\begin{figure}
	\centerline{\includegraphics[width=70mm]{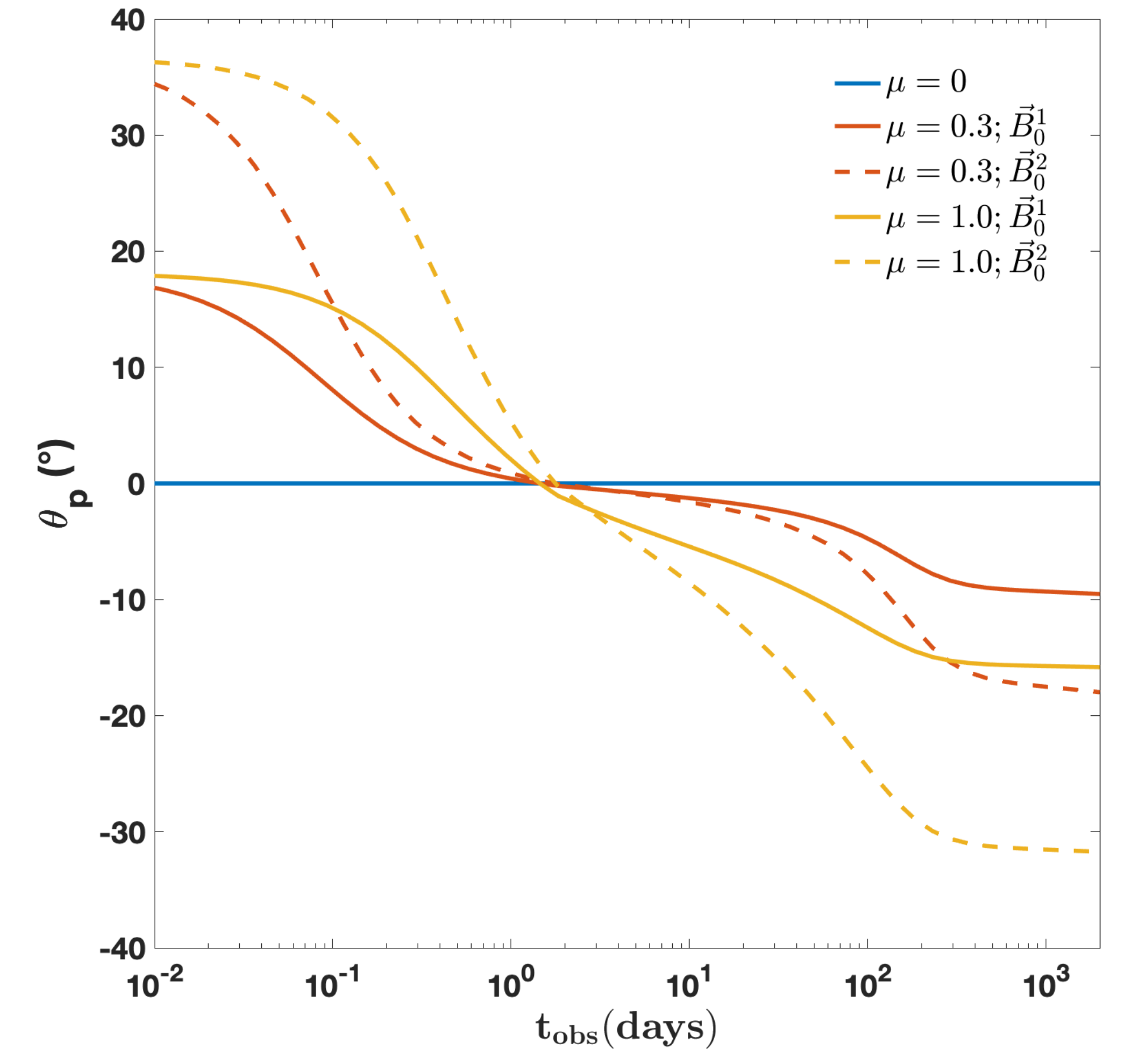}}
    \caption{Evolution of the polarization angle for different magnetic field configurations: a random field only, both ordered and random fields with different ratios of ordered to random components, $\theta_\mathrm{obs} = 2\: \theta_0 $. The other parameters are the same as in fig.\ \ref{fig:LC}.}
   \label{fig:polarization_angle}
\end{figure}

Comparing fig.\ \ref {fig:random_figure} and fig.\ \ref {fig:Q_ordered}, we can see that in some cases it can be hard to distinguish between a configuration with solely a random component and one with both ordered and random components for an observer outside of the jet. We argue here that a smoking gun for the presence of an ordered component would be the evolution of the polarization angle $\theta_p =0.5 \arctan (U/Q)$. Indeed, as seen previously, with only a random component, even anisotropic, the polarization has only two possible directions: along the $\bf{{\tilde{x}}}$ and $\bf{y}$ axes. However in the presence of an ordered component, even significantly smaller than the random one, the direction of polarization depends on the projection of the ordered component of the magnetic field, making many directions possible for the polarization vector.  In fig.\ \ref{fig:polarization_angle}, we can see the evolution of $\theta_p$ for different magnetic field configurations, for an observer outside the jet. In the presence of an ordered field, even significantly weaker than the random one, $\theta_p$ evolves slowly throughout the afterglow duration. It begins with a value which depends on the ordered component direction, and it decreases to almost the opposite value at late times. The angle is measured in the plane of the sky ($\bf{{\tilde{x}},y}$), therefore $\theta_p =0$ corresponds to a polarization along the $\bf{\tilde{x}}$ direction, the direction perpendicular to ${\bf n} \land {\bf n}_\mathrm{shock} $. We should note that if prior to compression, the ordered component lied in the plane of the observer, there is no break of symmetry and therefore $U=0$ and the polarization angle remain constant.

\begin{figure}
	\centerline{\includegraphics[width=80mm]{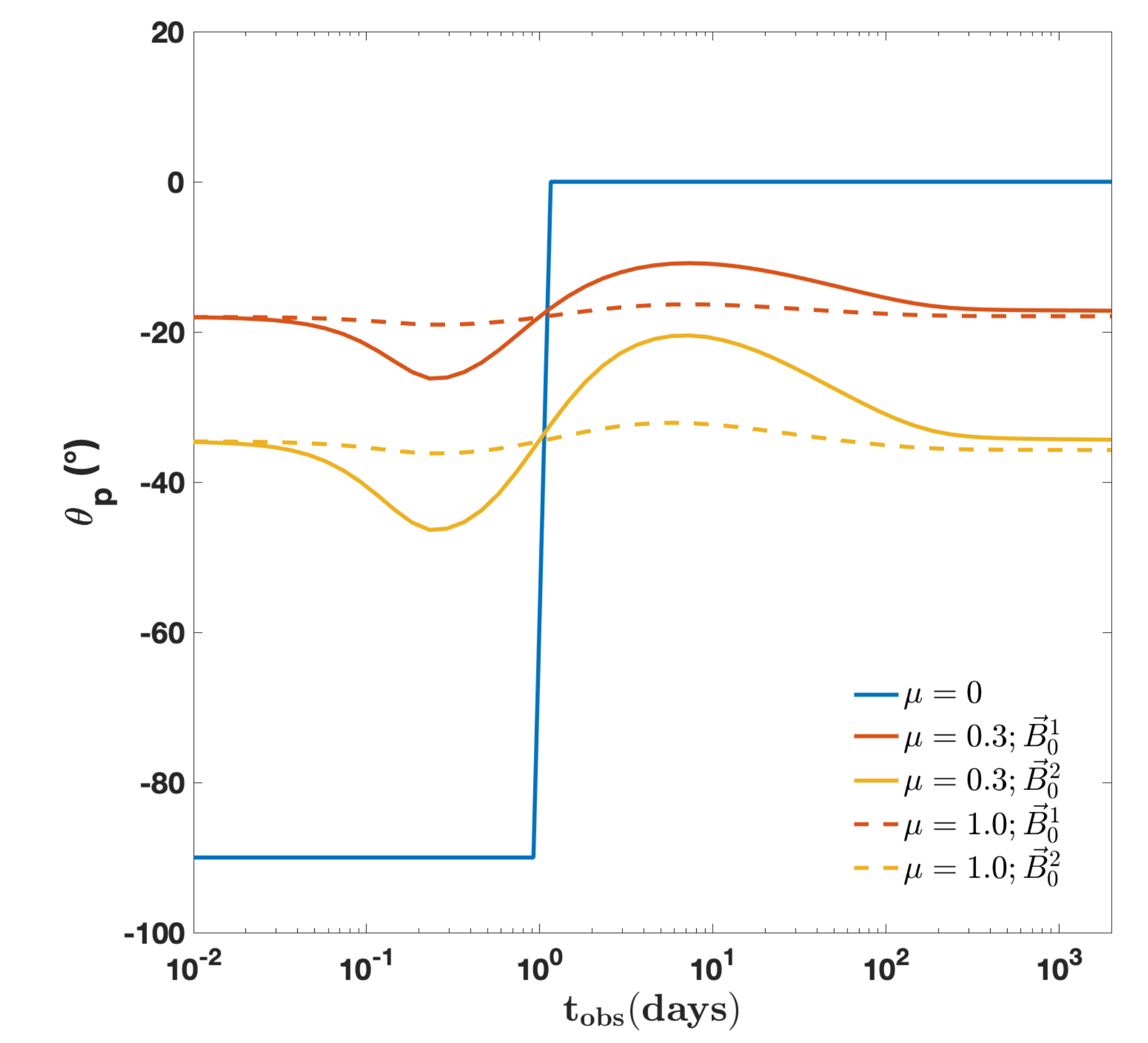}}
    \caption{Evolution of the polarization angle for different magnetic field configurations: random field only, both ordered and random fields with different ratios of ordered to random components, $\theta_\mathrm{obs} = 0.5\: \theta_0 $. The other parameters are the same as in fig.\ \ref{fig:LC}.}
   \label{fig:polarization_angle_inside}
\end{figure}

Depicted in fig.\ \ref {fig:polarization_angle_inside} is the evolution of $\theta_p$ for different magnetic field configurations, for an observer inside the jet. Without an ordered component, there is a 90\textdegree\ change of direction, from $\bf{y}$ to $\bf{\tilde{x}}$.  With an ordered component, there are either small variations (for a weaker ordered component) to no variation at all (for an equal ordered component). 

\section{Application to GRB170817}
\label{sec:GRB170817}

GW170817, the first gravitational waves (GW) signal from a binary neutron star merger was detected by advanced LIGO/Virgo on August 17 2017 \citep{Abbott}. It was accompanied by the first electromagnetic counterpart to any GW detection, the short gamma-ray burst, GRB 170817A  \citep{goldstein}. As this event was the first of its kind,  its afterglow was extensively monitored at all wavelengths (e.g. \cite{Alexander:18,davanzo,dobie,Hallinan:17,Lyman,makhathini,Margutti:17,Margutti:18}; ,\cite{Mooley:18a,Mooley:18b,Mooley:18c,nynka,Troja:18,Troja:20}) and an upper limit on polarization 244 days after the merger was found \citep{corsi}. The many detections allow us to have good constraints on many usually unknown parameters such as the observer angle, the half opening angle of the jet and the density, making it an ideal case to investigate the possible magnetic field configurations. 

The afterglow had an unusual rising which was argued to come from either a structured jet {\color{black}(e.g.} \cite{lamb,lazzati,Margutti:18,davanzo,res,troja:17}) or a quasi-isotropic (cocoon-dominated) outflow {\color{black} (e.g.} \cite{kasliwal,Mooley:18a}). More recently, the VLBI and VLA observations revealed that GW170817 involved a narrow jet $\theta_j 	\leq 5$\textdegree\  that dominated the late-time afterglow \citep{Mooley:18b,ghirlanda}. This implies that at the time of the afterglow peak and afterwards, the observed signal behaved like an afterglow of a top-hat jet seen at $\theta_\mathrm{obs} \gg \theta_j$. 

\begin{figure}
	\centerline{\includegraphics[width=85mm]{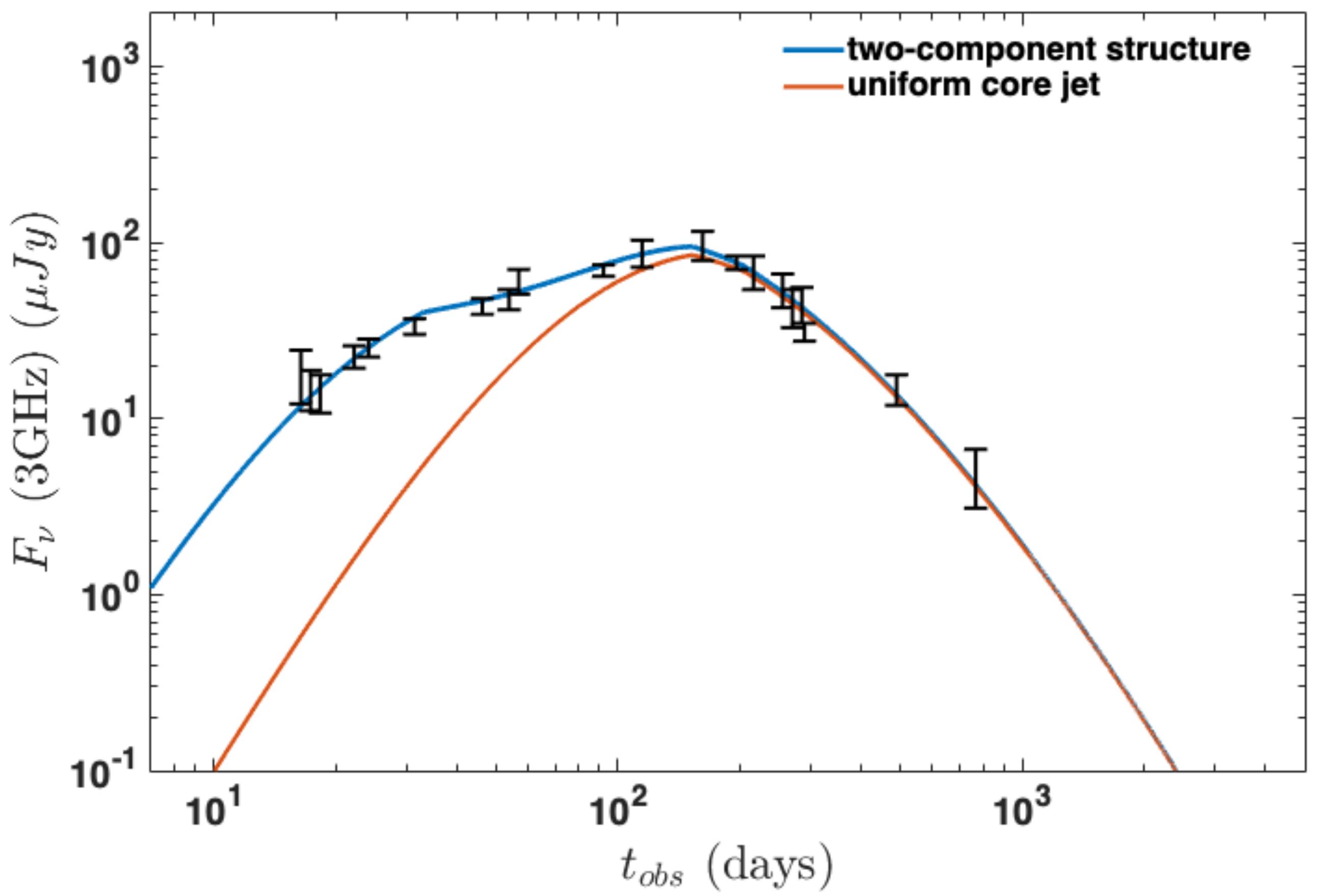}}
     \caption{ The $\nu$ = 3GHz light curves obtained with our semi-analytical code for both the two-component structure and the uniform core only. The parameters are $n=8.10^{-4}cm^{-3}, \epsilon_e = 0.01 , \epsilon_B= 0.001$,  {\color{black}$\theta_{obs}=22$}\textdegree\ , p=2.16. The error bars are the observed data points GW170817 from \citep{Hallinan:17,Alexander:17,Alexander:18,dobie,Margutti:18,Mooley:18a,Mooley:18c,makhathini}.}
   \label{fig:GRB17_LC}
\end{figure}

We used our semi analytical model to find a jet configuration that could i) reproduce the observations at all times including the unusual rise at the beginning, ii) whose core could account for the observations at the time of the peak and afterwards, iii) whose parameters are compatible with observation constraints. Our jet configuration is a two-component structure {\color{black}\footnote[6]{$\epsilon(\theta)=\epsilon_0 l(\theta)$ with $l(\theta)=1$ for $\theta \leq \theta_c$ and $l(\theta)=0.1$ for $\theta_c < \theta \leq \theta_0$}} consisting of a narrow uniform relativistic core with a half opening angle $\theta_0 = 3$\textdegree\ and an equivalent isotropic energy $10^{52}$erg, surrounded by wider mildly relativistic wings with 10$\% $ of the core energy. As seen in fig.\ \ref{fig:GRB17_LC}, our two-component structure fits all the data and considering only its uniform core, we can reproduce the afterglow at the time of the peak and afterwards. Moreover with our semi-analytical model we can easily check the Lorentz factor. At the time of the peak, $t \approx $ 155 days, we have $\Gamma \simeq 3.6 $, which is in good agreement with the observed Lorentz factor $\Gamma 	\simeq 4 \pm 0.5 $ \citep{Mooley:18b}. We then use the core jet parameters that fit the observations to investigate the magnetic field configuration in the vicinity of GRB170817 compatible with the upper limit polarization $\Pi< 12\% $, measured by \cite{corsi} at $t_\mathrm{obs}=244$ days and $\nu$=2.8 GHz. 
 
\begin{figure}
	\centerline{\includegraphics[width=90mm]{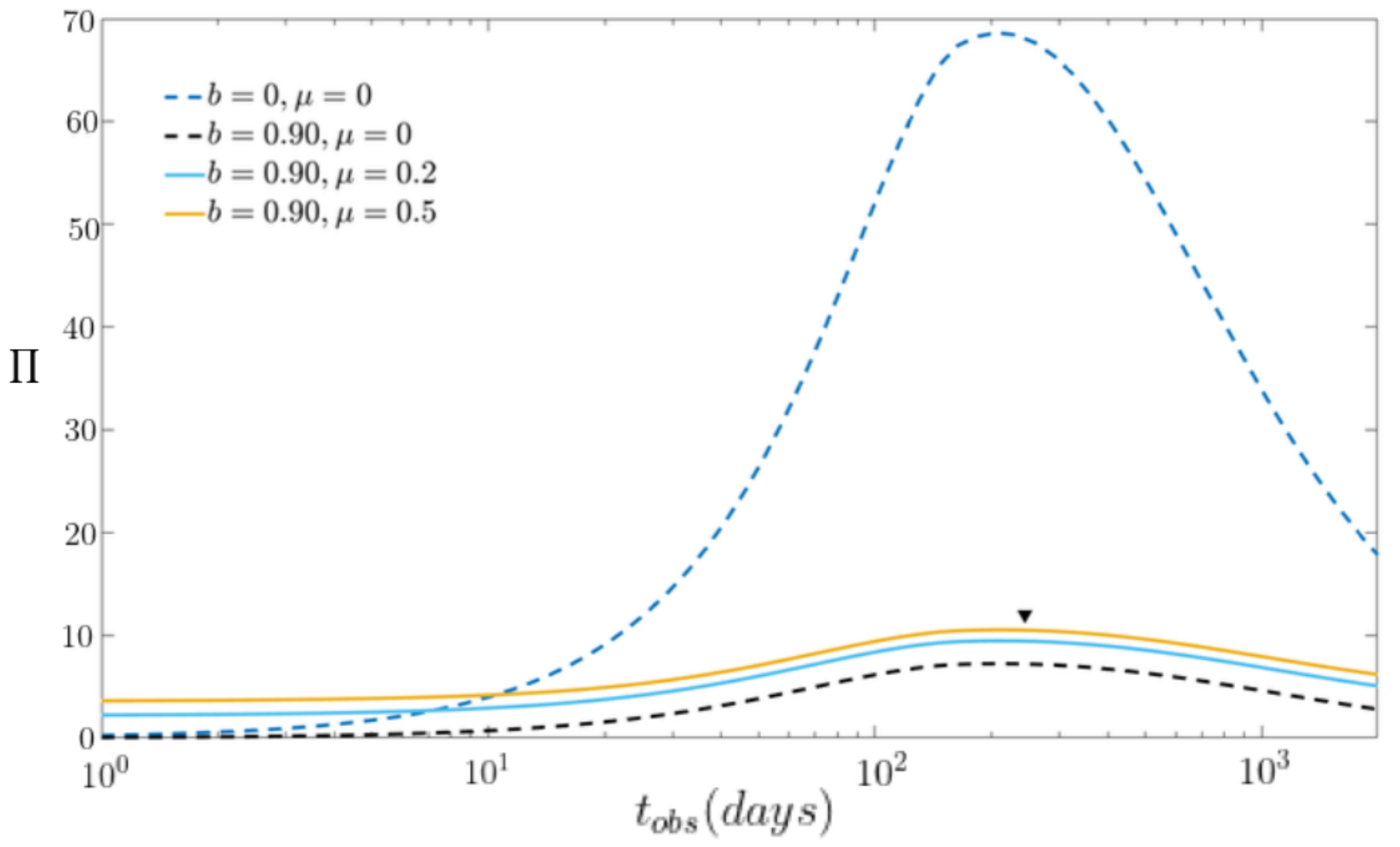}}
     \caption{ Polarization curves for different magnetic field configurations: random confined to the shock plane and random only with anisotropy in dashed lines, an anisotropic random component combined with an ordered component in full lines, the parameters are the same as in fig.\ref{fig:GRB17_LC}.}
   \label{fig:pola_GRB17}
\end{figure}

We tested different configurations of magnetic field and found that a random field confined to the shock plane is ruled out, see fig. \ref{fig:pola_GRB17}. However, a random field with an anisotropy factor b such as $ 0.85 \leq b \leq 1.18$ is compatible with the polarization upper limit (b=1 meaning isotropy). {\color{black} The  requirement for an almost isotropic random component was also derived independently in \cite{GG} }.  Moreover, we found that a magnetic field consisting of both random and a ordered components with $ 0.85 \leq b \leq 1.18$ and $\mu\leq0.5 $ is also compatible with the detection. Therefore, the polarization upper limit for GRB170817 requires an almost isotropic random component  and is compatible with the presence of an ordered component as large as 1/2  the random one.  
 
\section{Conclusions \&  Discussions}
\label{sec:conclusions}
In this work, we have studied the impact of different magnetic field configurations on afterglow polarization. We considered different plausible configurations including the one consisting of both ordered and random components which is the most probable if the compressed ISM magnetic field plays a role, even small. We found that the evolution of the polarization angle $\theta_p$ would be a smoking gun to confirm the presence of an ordered component, even significantly weaker than the random one. Indeed, for an observer inside of the jet, $\theta_p$ has small to no variation at all with an ordered component, instead of the 90\textdegree\ swing obtained with random component solely. For an observer outside of the jet, $\theta_p$ evolves slowly throughout the afterglow in the presence of an ordered component while there is no variation with a random component solely. For an observer inside the jet, polarization detections at all times can confirm or refute the presence of an ordered component and provide an indication to the ratio between the ordered and random components. For an observer outside the jet, we found that a polarization outside of the peak region would point towards the presence/absence of an ordered component and to the ratio between the ordered and random components. However in the region close to the peak, the polarization is a good indicator for the anisotropy factor. This was confirmed by the upper limit on polarization found for GRB170817 which was detected 244 days after the merger in the region close to the peak flux $t\approx $155 days and for which we could derive good constraints on the anisotropy factor \citep{corsi}. 

More specifically, for GRB170817 we found that the upper limit on polarization was compatible with an ordered component as large as half the random one and required a random component not confined to the shock plane. We obtained that the anisotropy factor should satisfy  $0.85\leq b \leq 1.18$ (with $b=1$ corresponding to isotropy, and $b=2$ corresponding to $\langle B_{\perp}\rangle = \langle B_{\parallel}\rangle $). The origin of such a parallel random component is puzzling. Indeed, one of the most popular scenario for shock generated magnetic field, that of  Weibel instability, would produce an almost completely transverse magnetic field \citep{lemoinegremillet,rashid}. Another popular scenario is the coherent patch scenario where shock generated magnetic field is ordered inside small patches, and patches ordered directions are incoherent between them \citep{gruzinovwaxman}. In such a scenario, the polarization is $\Pi \approx 70\% / \sqrt{N} $, with $N$ being the number of patches. Thus, the upper limit detection for GRB170817 requires $N\geq 35$. We can also calculate the size of the patches $ \theta_B$  using $N \approx ( \Gamma \theta_B)^{-2} $, which gives us  $ \theta_B \approx 0.053$ rad for $\Gamma \approx 3.5 $, deriving it from the observed $\Gamma\approx 4 $ at the time of the peak \citep{Mooley:18b}. However, observations found that the half opening angle was   $\theta_0 \leq $5\textdegree  = 0.087 rad  \citep{Mooley:18b} which seems to disfavour this scenario as well. Another possible scenario is the amplification of an existing magnetic field through MHD turbulence. Indeed, if the shock sweeps an inhomogeneous external medium, multiple vorticities arise downstream of the shock due to the growth of the Richtmyer-Meshkov instability, increasing the field strength \citep{sironigoodman,inoue,mizuno}. \cite{mizuno} and \cite{inoue} found a nearly isotropic turbulent density which seems to be favoured for GRB170817.

As showed in this work, polarization evolution highly depends on whether or not the observer is outside the jet. Therefore, future afterglow polarization detections with this information will allow us to find the magnetic field configurations compatible with those detections. If polarization detection or polarization angle requires an ordered component, it will confirm the importance of the compressed ISM magnetic field. Indeed, shock generated magnetic fields have a very small coherence length, of the order of the plasma skin depth. More polarization detections in the region of the peak will help us constrain the anisotropy factor and better understand the shock generated magnetic fields and therefore the collisionless shocks.  

\section*{Acknowledgements}

O.T would like to thank R. Moharana, N.Globus, J. Freundlich and N. Cornuault for our fruitful discussions. O.T is also grateful for the Einstein-Kaye scholarship. 

\section*{Data availability}
The data underlying this article will be shared on reasonable request to the corresponding author.



\bibliographystyle{mnras}
\bibliography{export-bibtex-3.bib,export-bibtex-4.bib,export-bibtex-5.bib,export-bibtex-9.bib} 

\begin{thebibliography}{}
\makeatletter
\relax
\def\mn@urlcharsother{\let\do\@makeother \do\$\do\&\do\#\do\^\do\_\do\%\do\~}
\def\mn@doi{\begingroup\mn@urlcharsother \@ifnextchar [ {\mn@doi@}
  {\mn@doi@[]}}
\def\mn@doi@[#1]#2{\def\@tempa{#1}\ifx\@tempa\@empty \href
  {http://dx.doi.org/#2} {doi:#2}\else \href {http://dx.doi.org/#2} {#1}\fi
  \endgroup}
\def\mn@eprint#1#2{\mn@eprint@#1:#2::\@nil}
\def\mn@eprint@arXiv#1{\href {http://arxiv.org/abs/#1} {{\tt arXiv:#1}}}
\def\mn@eprint@dblp#1{\href {http://dblp.uni-trier.de/rec/bibtex/#1.xml}
  {dblp:#1}}
\def\mn@eprint@#1:#2:#3:#4\@nil{\def\@tempa {#1}\def\@tempb {#2}\def\@tempc
  {#3}\ifx \@tempc \@empty \let \@tempc \@tempb \let \@tempb \@tempa \fi \ifx
  \@tempb \@empty \def\@tempb {arXiv}\fi \@ifundefined
  {mn@eprint@\@tempb}{\@tempb:\@tempc}{\expandafter \expandafter \csname
  mn@eprint@\@tempb\endcsname \expandafter{\@tempc}}}

\bibitem[\protect\citeauthoryear{{Abbott} et~al.,}{{Abbott}
  et~al.}{2017}]{Abbott}
{Abbott} B.~P.,  et~al., 2017, \mn@doi [\prl] {10.1103/PhysRevLett.119.161101},
  \href {https://ui.adsabs.harvard.edu/abs/2017PhRvL.119p1101A} {119, 161101}

\bibitem[\protect\citeauthoryear{{Alexander} et~al.,}{{Alexander}
  et~al.}{2017}]{Alexander:17}
{Alexander} K.~D.,  et~al., 2017, \mn@doi [\apjl] {10.3847/2041-8213/aa905d},
  \href {https://ui.adsabs.harvard.edu/abs/2017ApJ...848L..21A} {848, L21}

\bibitem[\protect\citeauthoryear{{Alexander} et~al.,}{{Alexander}
  et~al.}{2018}]{Alexander:18}
{Alexander} K.~D.,  et~al., 2018, \mn@doi [\apjl] {10.3847/2041-8213/aad637},
  \href {https://ui.adsabs.harvard.edu/abs/2018ApJ...863L..18A} {863, L18}

\bibitem[\protect\citeauthoryear{{Barniol Duran} \& {Kumar}}{{Barniol Duran} \&
  {Kumar}}{2011}]{barniol}
{Barniol Duran} R.,  {Kumar} P.,  2011, \mn@doi [\mnras]
  {10.1111/j.1365-2966.2011.19369.x}, \href
  {https://ui.adsabs.harvard.edu/abs/2011MNRAS.417.1584B} {417, 1584}

\bibitem[\protect\citeauthoryear{{Blandford} \& {McKee}}{{Blandford} \&
  {McKee}}{1976}]{blandford}
{Blandford} R.~D.,  {McKee} C.~F.,  1976, \mn@doi [Physics of Fluids]
  {10.1063/1.861619}, \href
  {https://ui.adsabs.harvard.edu/abs/1976PhFl...19.1130B} {19, 1130}

\bibitem[\protect\citeauthoryear{{Boulanger} et~al.,}{{Boulanger}
  et~al.}{2018}]{boulanger}
{Boulanger} F.,  et~al., 2018, \mn@doi [\jcap] {10.1088/1475-7516/2018/08/049},
  \href {https://ui.adsabs.harvard.edu/abs/2018JCAP...08..049B} {2018, 049}

\bibitem[\protect\citeauthoryear{{Corsi} et~al.,}{{Corsi} et~al.}{2018}]{corsi}
{Corsi} A.,  et~al., 2018, \mn@doi [\apjl] {10.3847/2041-8213/aacdfd}, \href
  {https://ui.adsabs.harvard.edu/abs/2018ApJ...861L..10C} {861, L10}

\bibitem[\protect\citeauthoryear{{Covino} \& {Gotz}}{{Covino} \&
  {Gotz}}{2016}]{Covino16}
{Covino} S.,  {Gotz} D.,  2016, Astronomical and Astrophysical Transactions,
  \href {https://ui.adsabs.harvard.edu/abs/2016A&AT...29..205C} {29, 205}

\bibitem[\protect\citeauthoryear{{Covino} et~al.,}{{Covino}
  et~al.}{1999}]{Covino99}
{Covino} S.,  et~al., 1999, \aap, \href
  {https://ui.adsabs.harvard.edu/abs/1999A&A...348L...1C} {348, L1}

\bibitem[\protect\citeauthoryear{{Covino}, {Ghisellini}, {Lazzati}  \&
  {Malesani}}{{Covino} et~al.}{2004}]{Covino04}
{Covino} S.,  {Ghisellini} G.,  {Lazzati} D.,   {Malesani} D.,  2004, in
  {Feroci} M.,  {Frontera} F.,  {Masetti} N.,   {Piro} L.,  eds,  Astronomical
  Society of the Pacific Conference Series Vol. 312, Gamma-Ray Bursts in the
  Afterglow Era. p.~169 (\mn@eprint {arXiv} {astro-ph/0301608})

\bibitem[\protect\citeauthoryear{{D'Avanzo} et~al.,}{{D'Avanzo}
  et~al.}{2018}]{davanzo}
{D'Avanzo} P.,  et~al., 2018, \mn@doi [\aap] {10.1051/0004-6361/201832664},
  \href {https://ui.adsabs.harvard.edu/abs/2018A&A...613L...1D} {613, L1}

\bibitem[\protect\citeauthoryear{{Dobie} et~al.,}{{Dobie} et~al.}{2018}]{dobie}
{Dobie} D.,  et~al., 2018, \mn@doi [\apjl] {10.3847/2041-8213/aac105}, \href
  {https://ui.adsabs.harvard.edu/abs/2018ApJ...858L..15D} {858, L15}

\bibitem[\protect\citeauthoryear{{Fong}, {Berger}, {Margutti}  \&
  {Zauderer}}{{Fong} et~al.}{2015}]{fong}
{Fong} W.,  {Berger} E.,  {Margutti} R.,   {Zauderer} B.~A.,  2015, \mn@doi
  [\apj] {10.1088/0004-637X/815/2/102}, \href
  {https://ui.adsabs.harvard.edu/abs/2015ApJ...815..102F} {815, 102}

\bibitem[\protect\citeauthoryear{{Ghirlanda} et~al.,}{{Ghirlanda}
  et~al.}{2019}]{ghirlanda}
{Ghirlanda} G.,  et~al., 2019, \mn@doi [Science] {10.1126/science.aau8815},
  \href {https://ui.adsabs.harvard.edu/abs/2019Sci...363..968G} {363, 968}

\bibitem[\protect\citeauthoryear{{Ghisellini} \& {Lazzati}}{{Ghisellini} \&
  {Lazzati}}{1999}]{Ghisellini:99}
{Ghisellini} G.,  {Lazzati} D.,  1999, \mn@doi [\mnras]
  {10.1046/j.1365-8711.1999.03025.x}, \href
  {https://ui.adsabs.harvard.edu/abs/1999MNRAS.309L...7G} {309, L7}

\bibitem[\protect\citeauthoryear{{Gill} \& {Granot}}{{Gill} \&
  {Granot}}{2018}]{Gill:18}
{Gill} R.,  {Granot} J.,  2018, \mn@doi [\mnras] {10.1093/mnras/sty1214}, \href
  {https://ui.adsabs.harvard.edu/abs/2018MNRAS.478.4128G} {478, 4128}

\bibitem[\protect\citeauthoryear{{Gill} \& {Granot}}{{Gill} \&
  {Granot}}{2020}]{GG}
{Gill} R.,  {Granot} J.,  2020, \mn@doi [\mnras] {10.1093/mnras/stz3340}, \href
  {https://ui.adsabs.harvard.edu/abs/2020MNRAS.491.5815G} {491, 5815}

\bibitem[\protect\citeauthoryear{{Goldstein} et~al.,}{{Goldstein}
  et~al.}{2017}]{goldstein}
{Goldstein} A.,  et~al., 2017, \mn@doi [\apjl] {10.3847/2041-8213/aa8f41},
  \href {https://ui.adsabs.harvard.edu/abs/2017ApJ...848L..14G} {848, L14}

\bibitem[\protect\citeauthoryear{{Granot} \& {K{\"o}nigl}}{{Granot} \&
  {K{\"o}nigl}}{2003}]{Granot:03}
{Granot} J.,  {K{\"o}nigl} A.,  2003, \mn@doi [\apjl] {10.1086/378733}, \href
  {https://ui.adsabs.harvard.edu/abs/2003ApJ...594L..83G} {594, L83}

\bibitem[\protect\citeauthoryear{{Granot} \& {Sari}}{{Granot} \&
  {Sari}}{2002}]{GS}
{Granot} J.,  {Sari} R.,  2002, \mn@doi [\apj] {10.1086/338966}, \href
  {https://ui.adsabs.harvard.edu/abs/2002ApJ...568..820G} {568, 820}

\bibitem[\protect\citeauthoryear{{Granot}, {Piran}  \& {Sari}}{{Granot}
  et~al.}{1999}]{GPS:99}
{Granot} J.,  {Piran} T.,   {Sari} R.,  1999, \mn@doi [\apj] {10.1086/306884},
  \href {https://ui.adsabs.harvard.edu/abs/1999ApJ...513..679G} {513, 679}

\bibitem[\protect\citeauthoryear{{Gruzinov}}{{Gruzinov}}{1999}]{gruzinov}
{Gruzinov} A.,  1999, \mn@doi [\apjl] {10.1086/312323}, \href
  {https://ui.adsabs.harvard.edu/abs/1999ApJ...525L..29G} {525, L29}

\bibitem[\protect\citeauthoryear{{Gruzinov} \& {Waxman}}{{Gruzinov} \&
  {Waxman}}{1999}]{gruzinovwaxman}
{Gruzinov} A.,  {Waxman} E.,  1999, \mn@doi [\apj] {10.1086/306720}, \href
  {https://ui.adsabs.harvard.edu/abs/1999ApJ...511..852G} {511, 852}

\bibitem[\protect\citeauthoryear{{Hallinan} et~al.,}{{Hallinan}
  et~al.}{2017}]{Hallinan:17}
{Hallinan} G.,  et~al., 2017, \mn@doi [Science] {10.1126/science.aap9855},
  \href {https://ui.adsabs.harvard.edu/abs/2017Sci...358.1579H} {358, 1579}

\bibitem[\protect\citeauthoryear{{He}, {Wu}, {Toma}, {Wang}  \&
  {M{\'e}sz{\'a}ros}}{{He} et~al.}{2011}]{he}
{He} H.-N.,  {Wu} X.-F.,  {Toma} K.,  {Wang} X.-Y.,   {M{\'e}sz{\'a}ros} P.,
  2011, \mn@doi [\apj] {10.1088/0004-637X/733/1/22}, \href
  {https://ui.adsabs.harvard.edu/abs/2011ApJ...733...22H} {733, 22}

\bibitem[\protect\citeauthoryear{{Inoue}, {Asano}  \& {Ioka}}{{Inoue}
  et~al.}{2011}]{inoue}
{Inoue} T.,  {Asano} K.,   {Ioka} K.,  2011, \mn@doi [\apj]
  {10.1088/0004-637X/734/2/77}, \href
  {https://ui.adsabs.harvard.edu/abs/2011ApJ...734...77I} {734, 77}

\bibitem[\protect\citeauthoryear{{Jordana-Mitjans} et~al.,}{{Jordana-Mitjans}
  et~al.}{2020}]{jordana}
{Jordana-Mitjans} N.,  et~al., 2020, \mn@doi [\apj] {10.3847/1538-4357/ab7248},
  \href {https://ui.adsabs.harvard.edu/abs/2020ApJ...892...97J} {892, 97}

\bibitem[\protect\citeauthoryear{{Kasliwal} et~al.,}{{Kasliwal}
  et~al.}{2017}]{kasliwal}
{Kasliwal} M.~M.,  et~al., 2017, \mn@doi [Science] {10.1126/science.aap9455},
  \href {https://ui.adsabs.harvard.edu/abs/2017Sci...358.1559K} {358, 1559}

\bibitem[\protect\citeauthoryear{{Kobayashi}}{{Kobayashi}}{2000}]{kobayashi}
{Kobayashi} S.,  2000, \mn@doi [\apj] {10.1086/317869}, \href
  {https://ui.adsabs.harvard.edu/abs/2000ApJ...545..807K} {545, 807}

\bibitem[\protect\citeauthoryear{{Kumar} \& {Barniol Duran}}{{Kumar} \&
  {Barniol Duran}}{2010}]{kumar}
{Kumar} P.,  {Barniol Duran} R.,  2010, \mn@doi [\mnras]
  {10.1111/j.1365-2966.2010.17274.x}, \href
  {https://ui.adsabs.harvard.edu/abs/2010MNRAS.409..226K} {409, 226}

\bibitem[\protect\citeauthoryear{{Laing}}{{Laing}}{1980}]{laing}
{Laing} R.~A.,  1980, \mn@doi [\mnras] {10.1093/mnras/193.3.439}, \href
  {https://ui.adsabs.harvard.edu/abs/1980MNRAS.193..439L} {193, 439}

\bibitem[\protect\citeauthoryear{{Lamb} \& {Kobayashi}}{{Lamb} \&
  {Kobayashi}}{2018}]{lamb}
{Lamb} G.~P.,  {Kobayashi} S.,  2018, \mn@doi [\mnras] {10.1093/mnras/sty1108},
  \href {https://ui.adsabs.harvard.edu/abs/2018MNRAS.478..733L} {478, 733}

\bibitem[\protect\citeauthoryear{{Laskar} et~al.,}{{Laskar}
  et~al.}{2019}]{laskar}
{Laskar} T.,  et~al., 2019, \mn@doi [\apjl] {10.3847/2041-8213/ab2247}, \href
  {https://ui.adsabs.harvard.edu/abs/2019ApJ...878L..26L} {878, L26}

\bibitem[\protect\citeauthoryear{{Lazzati}, {Perna}, {Morsony}, {Lopez-Camara},
  {Cantiello}, {Ciolfi}, {Giacomazzo}  \& {Workman}}{{Lazzati}
  et~al.}{2018}]{lazzati}
{Lazzati} D.,  {Perna} R.,  {Morsony} B.~J.,  {Lopez-Camara} D.,  {Cantiello}
  M.,  {Ciolfi} R.,  {Giacomazzo} B.,   {Workman} J.~C.,  2018, \mn@doi [\prl]
  {10.1103/PhysRevLett.120.241103}, \href
  {https://ui.adsabs.harvard.edu/abs/2018PhRvL.120x1103L} {120, 241103}

\bibitem[\protect\citeauthoryear{{Lemoine}, {Gremillet}, {Pelletier}  \&
  {Vanthieghem}}{{Lemoine} et~al.}{2019}]{lemoinegremillet}
{Lemoine} M.,  {Gremillet} L.,  {Pelletier} G.,   {Vanthieghem} A.,  2019,
  \mn@doi [\prl] {10.1103/PhysRevLett.123.035101}, \href
  {https://ui.adsabs.harvard.edu/abs/2019PhRvL.123c5101L} {123, 035101}

\bibitem[\protect\citeauthoryear{{Lyman} et~al.,}{{Lyman} et~al.}{2018}]{Lyman}
{Lyman} J.~D.,  et~al., 2018, \mn@doi [Nature Astronomy]
  {10.1038/s41550-018-0511-3}, \href
  {https://ui.adsabs.harvard.edu/abs/2018NatAs...2..751L} {2, 751}

\bibitem[\protect\citeauthoryear{{Makhathini} et~al.,}{{Makhathini}
  et~al.}{2020}]{makhathini}
{Makhathini} S.,  et~al., 2020, arXiv e-prints, \href
  {https://ui.adsabs.harvard.edu/abs/2020arXiv200602382M} {p. arXiv:2006.02382}

\bibitem[\protect\citeauthoryear{{Margutti} et~al.,}{{Margutti}
  et~al.}{2017}]{Margutti:17}
{Margutti} R.,  et~al., 2017, \mn@doi [\apjl] {10.3847/2041-8213/aa9057}, \href
  {https://ui.adsabs.harvard.edu/abs/2017ApJ...848L..20M} {848, L20}

\bibitem[\protect\citeauthoryear{{Margutti} et~al.,}{{Margutti}
  et~al.}{2018}]{Margutti:18}
{Margutti} R.,  et~al., 2018, \mn@doi [\apjl] {10.3847/2041-8213/aab2ad}, \href
  {https://ui.adsabs.harvard.edu/abs/2018ApJ...856L..18M} {856, L18}

\bibitem[\protect\citeauthoryear{{Medvedev} \& {Loeb}}{{Medvedev} \&
  {Loeb}}{1999}]{medvedev}
{Medvedev} M.~V.,  {Loeb} A.,  1999, \mn@doi [\apj] {10.1086/308038}, \href
  {https://ui.adsabs.harvard.edu/abs/1999ApJ...526..697M} {526, 697}

\bibitem[\protect\citeauthoryear{{Mizuno}, {Pohl}, {Niemiec}, {Zhang},
  {Nishikawa}  \& {Hardee}}{{Mizuno} et~al.}{2014}]{mizuno}
{Mizuno} Y.,  {Pohl} M.,  {Niemiec} J.,  {Zhang} B.,  {Nishikawa} K.-I.,
  {Hardee} P.~E.,  2014, \mn@doi [\mnras] {10.1093/mnras/stu196}, \href
  {https://ui.adsabs.harvard.edu/abs/2014MNRAS.439.3490M} {439, 3490}

\bibitem[\protect\citeauthoryear{{Mooley} et~al.,}{{Mooley}
  et~al.}{2018a}]{Mooley:18a}
{Mooley} K.~P.,  et~al., 2018a, \mn@doi [\nat] {10.1038/nature25452}, \href
  {https://ui.adsabs.harvard.edu/abs/2018Natur.554..207M} {554, 207}

\bibitem[\protect\citeauthoryear{{Mooley} et~al.,}{{Mooley}
  et~al.}{2018b}]{Mooley:18b}
{Mooley} K.~P.,  et~al., 2018b, \mn@doi [\nat] {10.1038/s41586-018-0486-3},
  \href {https://ui.adsabs.harvard.edu/abs/2018Natur.561..355M} {561, 355}

\bibitem[\protect\citeauthoryear{{Mooley} et~al.,}{{Mooley}
  et~al.}{2018c}]{Mooley:18c}
{Mooley} K.~P.,  et~al., 2018c, \mn@doi [\apjl] {10.3847/2041-8213/aaeda7},
  \href {https://ui.adsabs.harvard.edu/abs/2018ApJ...868L..11M} {868, L11}

\bibitem[\protect\citeauthoryear{{Nynka}, {Ruan}, {Haggard}  \&
  {Evans}}{{Nynka} et~al.}{2018}]{nynka}
{Nynka} M.,  {Ruan} J.~J.,  {Haggard} D.,   {Evans} P.~A.,  2018, \mn@doi
  [\apjl] {10.3847/2041-8213/aad32d}, \href
  {https://ui.adsabs.harvard.edu/abs/2018ApJ...862L..19N} {862, L19}

\bibitem[\protect\citeauthoryear{{Planck Collaboration} et~al.,}{{Planck
  Collaboration} et~al.}{2016}]{planckadam}
{Planck Collaboration} et~al., 2016, \mn@doi [\aap]
  {10.1051/0004-6361/201425044}, \href
  {https://ui.adsabs.harvard.edu/abs/2016A&A...586A.135P} {586, A135}

\bibitem[\protect\citeauthoryear{{Resmi} et~al.,}{{Resmi} et~al.}{2018}]{res}
{Resmi} L.,  et~al., 2018, \mn@doi [\apj] {10.3847/1538-4357/aae1a6}, \href
  {https://ui.adsabs.harvard.edu/abs/2018ApJ...867...57R} {867, 57}

\bibitem[\protect\citeauthoryear{{Rossi}, {Lazzati}, {Salmonson}  \&
  {Ghisellini}}{{Rossi} et~al.}{2004}]{rossi}
{Rossi} E.~M.,  {Lazzati} D.,  {Salmonson} J.~D.,   {Ghisellini} G.,  2004,
  \mn@doi [\mnras] {10.1111/j.1365-2966.2004.08165.x}, \href
  {https://ui.adsabs.harvard.edu/abs/2004MNRAS.354...86R} {354, 86}

\bibitem[\protect\citeauthoryear{{Santana}, {Barniol Duran}  \&
  {Kumar}}{{Santana} et~al.}{2014}]{santana}
{Santana} R.,  {Barniol Duran} R.,   {Kumar} P.,  2014, \mn@doi [\apj]
  {10.1088/0004-637X/785/1/29}, \href
  {https://ui.adsabs.harvard.edu/abs/2014ApJ...785...29S} {785, 29}

\bibitem[\protect\citeauthoryear{{Sari}}{{Sari}}{1999}]{Sari:99}
{Sari} R.,  1999, \mn@doi [\apjl] {10.1086/312294}, \href
  {https://ui.adsabs.harvard.edu/abs/1999ApJ...524L..43S} {524, L43}

\bibitem[\protect\citeauthoryear{{Sari} \& {Piran}}{{Sari} \&
  {Piran}}{1999}]{SP:99}
{Sari} R.,  {Piran} T.,  1999, \mn@doi [\apj] {10.1086/307508}, \href
  {https://ui.adsabs.harvard.edu/abs/1999ApJ...520..641S} {520, 641}

\bibitem[\protect\citeauthoryear{{Sedov}}{{Sedov}}{1959}]{Sedov}
{Sedov} L.~I.,  1959, {Similarity and Dimensional Methods in Mechanics}

\bibitem[\protect\citeauthoryear{{Shaisultanov}, {Lyubarsky}  \&
  {Eichler}}{{Shaisultanov} et~al.}{2012}]{rashid}
{Shaisultanov} R.,  {Lyubarsky} Y.,   {Eichler} D.,  2012, \mn@doi [\apj]
  {10.1088/0004-637X/744/2/182}, \href
  {https://ui.adsabs.harvard.edu/abs/2012ApJ...744..182S} {744, 182}

\bibitem[\protect\citeauthoryear{{Sironi} \& {Goodman}}{{Sironi} \&
  {Goodman}}{2007}]{sironigoodman}
{Sironi} L.,  {Goodman} J.,  2007, \mn@doi [\apj] {10.1086/523636}, \href
  {https://ui.adsabs.harvard.edu/abs/2007ApJ...671.1858S} {671, 1858}

\bibitem[\protect\citeauthoryear{{Taylor}}{{Taylor}}{1950}]{Taylor}
{Taylor} G.,  1950, \mn@doi [Proceedings of the Royal Society of London Series
  A] {10.1098/rspa.1950.0049}, \href
  {https://ui.adsabs.harvard.edu/abs/1950RSPSA.201..159T} {201, 159}

\bibitem[\protect\citeauthoryear{{Troja} et~al.,}{{Troja}
  et~al.}{2017}]{troja:17}
{Troja} E.,  et~al., 2017, \mn@doi [\nat] {10.1038/nature24290}, \href
  {https://ui.adsabs.harvard.edu/abs/2017Natur.551...71T} {551, 71}

\bibitem[\protect\citeauthoryear{{Troja} et~al.,}{{Troja}
  et~al.}{2018}]{Troja:18}
{Troja} E.,  et~al., 2018, \mn@doi [\mnras] {10.1093/mnrasl/sly061}, \href
  {https://ui.adsabs.harvard.edu/abs/2018MNRAS.478L..18T} {478, L18}

\bibitem[\protect\citeauthoryear{{Troja} et~al.,}{{Troja}
  et~al.}{2020}]{Troja:20}
{Troja} E.,  et~al., 2020, arXiv e-prints, \href
  {https://ui.adsabs.harvard.edu/abs/2020arXiv200601150T} {p. arXiv:2006.01150}

\bibitem[\protect\citeauthoryear{{Wijers} et~al.,}{{Wijers}
  et~al.}{1999}]{Wijers}
{Wijers} R.~A.~M.~J.,  et~al., 1999, \mn@doi [\apjl] {10.1086/312262}, \href
  {https://ui.adsabs.harvard.edu/abs/1999ApJ...523L..33W} {523, L33}

\bibitem[\protect\citeauthoryear{{Zhang} et~al.,}{{Zhang} et~al.}{2019}]{zhang}
{Zhang} S.-N.,  et~al., 2019, \mn@doi [Nature Astronomy]
  {10.1038/s41550-018-0664-0}, \href
  {https://ui.adsabs.harvard.edu/abs/2019NatAs...3..258Z} {3, 258}

\makeatother
\end{thebibliography}





\bsp	
\label{lastpage}
\end{document}